\documentclass[a4paper,fleqn,usenatbib,useAMS]{mnras}

\usepackage{graphicx}	
\usepackage{amsmath}	
\usepackage{amssymb}	
\usepackage{multicol}        
\usepackage{bm}		
\usepackage{pdflscape}	
\usepackage{multirow}
\usepackage{xcolor}
\usepackage{hyperref}

\newcommand{\tabincell}[2]{\begin{tabular}{@{}#1@{}}#2\end{tabular}}

\usepackage[T1]{fontenc}
\usepackage{ae,aecompl}

\usepackage{newtxtext,newtxmath}

\title[Galaxy groups at high redshift]{Identifying galaxy 
	groups at high redshift from incomplete 
	spectroscopic data: I. The group finder and application 
	to zCOSMOS}

\author[Wang et al.]{
	Kai Wang,$^{1,2}$\thanks{Contact e-mail: wkcosmology@gmail.com\href{}{}}\thanks{Present address: Department of Astronomy, University of Massachusetts Amherst, MA 01003, USA}
	H.J. Mo,$^{2}$
	Cheng Li,$^{1}$
	Jiacheng Meng$^{1}$
	and Yangyao Chen$^{1,2}$
	\\
	$^{1}$Department of Astronomy, Tsinghua University, Beijing 100084, China\\
	$^{2}$Department of Astronomy, University of Massachusetts Amherst, MA 01003, USA}

\date{Last updated 2019 May 22; in original form 2018 September 5}

\pubyear{2020}

\begin{document}
	\label{firstpage}
	\pagerange{\pageref{firstpage}--\pageref{lastpage}}
	\maketitle
	
\begin{abstract}
	Identifying galaxy groups from redshift surveys of galaxies
	plays an important role in 
	connecting galaxies with the underlying dark matter distribution. 
	Current and future high-$z$ spectroscopic surveys, usually 
	incomplete in redshift sampling, present both opportunities 
	and challenges to identifying groups in the high-$z$ Universe. 
    We develop a group finder that is based on incomplete redshift 
    samples combined with photometric data, using a machine 
    learning method to assign halo masses to identified groups.  
    Test using realistic mock catalogs shows that 
    $\gtrsim 90\%$ of true groups with halo masses 
    $\rm M_h \gtrsim 10^{12} M_{\odot}/h$ are successfully identified, 
    and that the fraction of contaminants is smaller than $10\%$. The standard 
    deviation in the halo mass estimation is smaller than 0.25 dex
    at all masses. 
    We apply our group finder to zCOSMOS-bright and describe basic 
    properties of the group catalog obtained.
\end{abstract}

\begin{keywords}
	methods: statistical - galaxies: groups: general - dark matter - large-scale structure of Universe
\end{keywords}
	
	
	
\section{Introduction}%
\label{sec:introduction}

Identifying galaxy groups/clusters from galaxy surveys is a
practice that can be dated back to \citet{abell1958distribution}, 
who identified 2,700 clusters from the Palomar Observatory Sky Survey 
(POSS) using the distribution of galaxies in the sky.  
Similar investigations have been carried out later 
by \citet{zwicky1966catalogue} and \citet{abell_catalog_1989}.  
Without distance information, these catalogs can be contaminated 
severely by projection effects. With the advent of large redshift 
surveys of galaxies, efforts have been made to identify galaxy 
clusters/groups (collectively referred to as galaxy groups in the 
following) using the galaxy distribution in redshift space. 
For example, galaxy groups have been identified from the CfA 
redshift survey \citep[e.g.][]{huchraGroupsGalaxiesNearby1982}, the Two Degree 
Field Galaxy Redshift Survey \citep[e.g.][]{Eke:2004bf, yang2005halo, tagoClustersGroupsGalaxies2006}, 
the Two Micron All Sky Redshift Survey 
\citep[e.g.][]{lavaux20112m++, tullyGALAXYGROUPS2MASS2015, crook2007groups}, and the
Sloan Digital Sky Survey \citep[e.g.][]{goto2005velocity,
berlind2006percolation, yang_galaxy_2007}.

In the contemporary paradigm of structure formation, the matter content 
of the universe is dominated by dark matter, and the structure in the cosmic 
density field forms hierarchically through gravitational instability. 
The virialized parts of the structure, commonly referred to as dark 
matter halos, are the places where galaxies form and evolve
\citep[see][for a review]{mo2010galaxy}. Since the relationship 
between the distribution of halos and the underlying density field 
is well understood \citep[e.g.][]{mo1996analytic}, one can use halos 
to trace the cosmic density field. Thus, there is a strong motivation 
to select galaxy groups to represent dark matter halos in the observed 
universe. With this in mind, many of the group catalogs 
published recently have been constructed using methods that are 
calibrated with galaxy occupations in dark matter halos 
\citep[e.g.][]{yang2005halo, yang_galaxy_2007, 2011arXiv1107.5046T, duarteMaggieModelsAlgorithms2015,
luGALAXYGROUPS2MASS2016, lim2017galaxy}. 

As virialized regions in the cosmic density field, galaxy 
groups can be used to investigate the role played 
by environment in galaxy formation and evolution, and 
a wealth of investigations have been carried out in this area. 
For example, the group-galaxy cross-correlation function 
can be used not only to probe how galaxies are distributed in 
halos but also to verify the presence of transition from 
the one-halo to two-halo terms \citep[e.g.][]{yangCrosscorrelationGalaxiesGroups2005, coilDEEP2GalaxyRedshift2006a, knobelGROUPGALAXYCROSSCORRELATIONFUNCTION2012}.
\citet{weinmann_properties_2006} studied the dependence of galaxy 
properties on their host halos, and found a strong correlation 
in the properties of galaxies residing in common dark matter halos, 
a phenomenon now referred to as {\it galactic conformity}
\citep[see also][]{knobelQUENCHINGSTARFORMATION2015, kawinwanichakijSATELLITEQUENCHINGGALACTIC2016, darvishCosmicWebGalaxies2017}.
\citet{wang2018elucid} found that the apparent dependence of the 
quenched fraction of galaxies on large-scale environment is 
largely induced by the dependence of quenching 
on the host halo mass combined with the biased distribution of 
dark matter halos in the cosmic density field. 
By stacking galaxy groups of similar mass, one can also 
extract the weak signal of Sunyaev-Zel'dovich (SZ) effects 
produced by the gas associated with dark matter halos over a large 
halo mass range \citep[e.g.][]{liProbingHotGas2011, 
vikramMeasurementGalaxyGroupThermal2017,limGasContentsGalaxy2018, 
lim2020detection}. A similar approach can also be applied to extract 
weak gravitational lensing signals produced by galaxy groups 
\citep[e.g.][]{mandelbaumDensityProfilesGalaxy2006, yangWeakLensingGalaxies2006, hanGalaxyMassAssembly2015, violaDarkMatterHalo2015, luo2018galaxy},
and to obtain the halo occupation distribution or conditional luminosity
functions of galaxies in halos of different masses 
\citep[e.g.][]{yang_galaxy_2008, yang_galaxy_2009,
rodriguezTakingAdvantagePhotometric2015, Lan:2016id}.

Since galaxy groups and the corresponding dark matter halos are 
biased tracers of the underlying density field, the group/halo population 
can also be used to reconstruct the current cosmic density 
field \citep{Wang_Mo_Jing_Guo_vandenBosch_Yang_2009, munoz-cuartasHalobasedReconstructionCosmic2011} and 
to constrain the initial conditions that produced the observed cosmic 
web \citep[e.g.][]{wangELUCIDEXPLORINGLOCAL2016}. 
Such reconstructions can not only help to quantify the mass density 
field within which real galaxies reside, but also provide information 
about the formation history of the observed cosmic web.  

So far galaxy group catalogs have been constructed mainly for the 
low-redshift Universe, where large and complete redshift surveys 
of galaxies are available. The situation is expected to change, 
as a number of large surveys of high-$z$ galaxies have been 
or are being carried out: for example 
VVDS \citep{LeFevre:2005bs},
ORELSE \citep{lubinOBSERVATIONSREDSHIFTEVOLUTION2009},
zCOSMOS \citep{lillyZCOSMOSLargeVLT2007}, 
DEEP2 \citep{Newman:2013ha}, 
VIPERS \citep{guzzoVIMOSPublicExtragalactic2014}, 
and PFS \citep{takadaExtragalacticScienceCosmology2014}.
However, surveys at high-$z$ are distinguished from their 
low-$z$ counterparts. Because of detection and time limits, 
redshift sampling in a high-$z$ survey is usually 
incomplete. For example, the zCOSMOS-bright survey
has a sampling rate of $\sim 55\%$ and the planed PFS a sampling 
rate of $\sim 70\%$.
The sampling rate may even be inhomogeneous across the sky -- for 
example, fiber collisions can make the sampling rate lower in higher 
density regions. In addition, since higher-$z$ galaxies are on average  
fainter, it is more difficult for a high-$z$ survey to include galaxies 
of low luminosities. Both of these make it more challenging to 
identify galaxy groups from high-$z$ data reliably.
Nevertheless, there have been attempts to identify galaxy 
groups from such incomplete spectroscopic samples
\citep[e.g.][]{knobel2009optical, Knobel:2012je, 2005ApJ...625....6G, gerkeIMPROVEDMOCKGALAXY2013, cucciatiVIMOSVLTDeepSurvey2010},
although one must be cautious about the uncertainties 
such incomplete sampling may generate.
On the other hand, almost all high-$z$ spectroscopic surveys are based on deep 
photometric surveys with multi-waveband information that 
can be used to obtain photometric redshifts as well as 
to estimate colors, luminosities and stellar masses of 
individual galaxies. This information can be combined with the spectroscopic 
data to improve group identifications. Indeed, such an approach 
has been applied in some previous investigations 
\citep[e.g.][]{Knobel:2012je}. There have also been attempts to 
identify galaxy groups using only photometric data 
\citep[e.g.][]{liFINDINGGALAXYGROUPS2008, gillisGroupfindingPhotometricRedshifts2011, oguri2018optically, euclidcollaborationEuclidPreparationIII2019, maturi_span_2019}.
The goal of this paper is to develop a group finding 
algorithm that is suitable for high-redshift surveys with incomplete redshift 
sampling. Our method combines spectroscopic galaxies with those
in the corresponding parent photometric survey 
to make full use of the information provided by galaxy 
clustering in the observational data. We aim to identify all groups
above a certain halo mass so as to obtain a complete group 
catalog to represent the dark matter halo population.  
We calibrate and test our group finder using detailed mock catalogs 
that mimick real observations at high redshift.
As an application, we apply our method to zCOSMOS-bright survey 
\citep[][]{lillyZCOSMOSLargeVLT2007, Lilly:2009fb}.

The structure of the paper is as follows. In \S\ref{sec:method}, we 
describe our group finding method, including identifications of groups 
from spectroscopic data and the incorporation of photometric galaxies. 
The mock catalogs used to test our group finder is presented 
in \S\ref{sec:mock_catalogs}. We test the performance of our group finding method, 
including halo mass assignment, in 
\S\,\ref{sec:testing_the_performances_of_the_group_finder}. 
The application of our method to the zCOSMOS-bright survey is presented 
in \S\ref{sec:the_application_to_the_zcosmos_sample}. 
Finally, we summarize our main results 
in \S\,\ref{sec:summary}.
Throughout the paper,  cosmological parameters 
are adopted from \citet{dunkleyFIVEYEARWILKINSONMICROWAVE2009}:
matter density parameter $\Omega_{\rm m}=0.258$, cosmological 
constant $\Omega_\Lambda=0.742$, reduced Hubble constant $h=0.72$, 
and primordial power index $n=0.96$.

\section{Method}%
\label{sec:method}

Different group finding methods have been proposed to identify galaxy groups 
from both spectroscopic and photometric surveys of galaxies, 
such as the Friends-of-Friends (FoF) grouping algorithm
\citep[e.g.][]{huchraGroupsGalaxiesNearby1982, davis1985evolution, Eke:2004bf, knobel2009optical}, 
the Voronoi-Delaunay Method 
\citep[VDM,][]{marinoniThreedimensionalIdentificationReconstruction2002a, 2005ApJ...625....6G, knobel2009optical}, 
the halo-based group finder \citep[e.g.][]{yang2005halo}, and the 
adaptive matched filter method \citep[e.g.][]{kepnerAutomatedClusterFinder1999, 2008ApJ...676..868D}.
In this paper, we will use a version of the FoF group 
finder to select potential groups, and test its performance 
for high-$z$ surveys where spectroscopic redshifts are usually 
incomplete\footnote{We note that methods,  
	such as the halo-based method and the matched filter method,   
	are not suitable for galaxy surveys with severe 
	redshift incompleteness, 
	because these methods need reliable halo mass estimates 
	to assign group memberships.}. 
After identifying potential groups with spectroscopic galaxies, we will 
examine how the inclusion of galaxies with photometric 
information can improve the quality of the selected  
groups in their ability of representing dark matter halos.

\subsection{The Friends-of-Friends method}%
\label{sub:the_friend_of_friend_method}

The FoF group finding algorithm is the simplest and 
one of the most commonly used method to identify galaxy 
groups from redshift surveys of galaxies 
\citep[e.g.][]{huchraGroupsGalaxiesNearby1982, davis1985evolution, Eke:2004bf, knobel2009optical}. 
The basic idea of this algorithm is to assign two galaxies into a 
common group if they satisfy the following criteria:
\begin{align}
    \theta_{ij} &\leqslant \frac12 \left(\frac{l_{\bot,i}}{d_{i}} + \frac{l_{\bot, j}}{d_{j}}\right)\\
    |d_i - d_j| & \leqslant \frac{l_{\Vert, i} + l_{\Vert, j}}{2}
\end{align}
where $\theta_{ij}$ is the angular separation of the two galaxies, 
$d_{i}$ and $d_{j}$ are their co-moving distances. The two length 
scales, $l_{\bot}$ and $l_{\Vert}$ in the above equations are defined as
\begin{align}
    l_{\bot, i} &= {\rm min}\left[l_{\rm max}(1+z_i),~\frac{b}
    {\bar n^{1/3}(\alpha_i, \delta_i, z_i)}\right]\\
    l_{\Vert, i} &= R\cdot l_{\bot, i},
\end{align}
where $b$ is the transverse linking length in units of the mean separation 
between galaxies, and $R$ is the ratio of the line-of-sight (los) 
linking length to the transverse one. To avoid the linking length 
from becoming unreasonably large in low density regions, $l_{\rm max}$ is 
employed to set a limit. In general, the sampling rate of galaxy redshift 
may change with both redshift and position in the sky (see below). 
We take into account the effect of such a sampling by using a local  
mean number density defined as 
\begin{equation}
    \bar n(\alpha, \delta, z) = \bar n(z) \times \frac{C(\alpha,\delta)}{\bar C}
    \label{eq_num_density}
\end{equation}
where $\bar n(z)$ is the number density of spectroscopic galaxies 
at redshift $z$. The completeness, $C(\alpha,\delta)$, is 
the number ratio between galaxies with spectroscopic redshift and 
all the galaxies that satisfy the sample selection criteria  
at a given sky position $(\alpha,\delta)$, and $\bar C$ is the 
number ratio of all the spectroscopic galaxies to all the galaxies 
satisfying the selection criteria. Altogether, the group finder contains 
three free parameters: $l_{\rm max}$, $b$ and $R$, which are tuned 
to achieve an optimal performance (see below).

\subsection{Supplementing with photometric galaxies}%
\label{sub:supplementing_with_photometric_galaxies}
	
Spectroscopic observations are usually shallower than the corresponding 
photometric catalogs from which targets for spectroscopic 
observation are selected, and different surveys usually have different 
target selection criteria. For high-$z$ spectroscopic surveys, 
a large fraction of the target galaxies may not have redshift 
measurements owing to observational limitations. 
Thus, the final product of a redshift survey depends 
both on its target selection criteria and its redshift sampling 
rate. In general, the incompleteness produced by these two 
factors depends not only on galaxy properties such as color, 
but also on the local number density of galaxies.
Because of this, the average sampling rate alone cannot 
characterize a survey completely.
Incomplete sampling introduces two problems for group identifications. 
First, a group may miss most of its member galaxies in the spectroscopic 
sample, especially for a poor system. Some groups may, therefore, 
be totally missed in the selection from the spectroscopic sample. 
Second, a group may miss its dominating member galaxy
(its central galaxy) in the spectroscopic data. In this case,
the group could be identified but its halo mass will be wrongly determined.

Meanwhile, high quality multi-wavelength photometric data are usually 
available not only for all target galaxies for spectroscopy,
but also for other galaxies down to a fainter magnitude.  
Such photometric data can be used not only to obtain sky positions and 
colors for these galaxies, but also to determine their photometric 
redshifts (photo-$z$), providing useful distance information. 
In particular, estimates of luminosity and stellar mass 
can be obtained from modeling the spectral energy distribution 
provided by the multi-wavelength photometric data for individual galaxies. 
All these can be used together with the spectroscopic data to 
improve group identifications.  
	
To tackle the two problems described above,
we focus on two populations of galaxies in the photometric sample. 
The first is {\it group central}, defined as the central galaxy of a group
whose members are correctly assigned to a galaxy group in the spectroscopic 
data. The second is {\it isolated central}, defined as a central galaxy
whose group members are completely missed in the spectroscopic sample.
We use information provided by all the spectroscopic groups around each 
photometric galaxy to determine the status of the galaxy.   
To do this, we select, for each photometric galaxy,  
$n_g$ closest (based on a projected distance, $r_p$) groups 
identified from the spectroscopic data that satisfy 
\begin{equation}
	\Delta z \leq 3\sigma_{z, \rm phot}(1 + z)
\end{equation}
where $\Delta z$ is the redshift difference between 
the photometric galaxy and the most massive galaxy in
the identified spectroscopic group,
and $\sigma_{z, \rm phot}$ is the uncertainty 
of the photo-$z$. 
The choice of $\Delta z$ is to ensure that most of the 
true centrals are included; the final choice is to be 
made by the machine learning algorithm described below.
The features to be used are quantities describing the 
relationship between the photometric galaxy and the $n_g$ spectroscopic 
groups, which are:
\begin{enumerate}
	\item $M_{*,\rm phot}$:
        the stellar mass of the photometric galaxy;
	\item $(r_{p,1}, r_{p,2}, ...r_{p,{n_g}})$:
        the projected distances between the photometric galaxy and the surrounding $n_g$ groups;
	\item $(\Delta z_1, \Delta z_2,... \Delta z_{n_g})$:
        the absolute value of redshift differences between the photometric galaxy and the surrounding $n_g$ groups;
	\item $(\Delta M_{*, 1}, \Delta M_{*, 2}, ...\Delta M_{*, n_g})$:
        the logarithm of the stellar mass ratio between the 
        photometric galaxy and the most massive galaxy of the surrounding $n_g$ groups.
\end{enumerate}
Thus, for each photometric galaxy, we have $3n_g +1$ features. 
The target is to describe the real relationship 
of the photometric galaxy with the $n_g$ surrounding groups. 
To this end, we define the target as a vector of $n_g + 1$ boolean values, with 
its first component indicating whether or not the photometric galaxy 
is a central, and the remaining $n_g$ components indicating if the galaxy  
belongs to group $i~(i=1,2,...n_g)$. 
	
We employ a powerful machine learning algorithm, the Random 
Forest Classifier (RFC) in \texttt{scikit-learn} \citep{scikit-learn}, 
to do the classification for photometric galaxies. We consider the 
photometric galaxy sample as a set of objects, 
\begin{equation}
	\mathcal{D}=\{ {\bm x}_i, {\bm y}_i \}_{i=1}^{|\mathcal{D}|},
	\quad ({\bm x}_i\in \mathcal{X},~~ { \bm y}_i\in \mathcal{Y})
\end{equation}
where ${\bm x}_i$ represents the features for the $i$-th photometric galaxy 
as listed above and is a point in the feature space $\mathcal{X}$, 
${\bm y}_i$ denotes the target vector defined above and is a point in 
the target space $\mathcal{Y}$, and $\mathcal{D}$ stands for the 
photometric galaxy sample with its size denoted by $|\mathcal{D}|$.	
The RFC is an ensemble of many decision trees, each of which  
is constructed from a bootstrap sample, 
$\mathcal{D}_{\rm bts}\in \mathcal{D}$, which is selected 
from the original sample $\mathcal{D}$ and only retains 
a randomly-chosen subset of the features for individual galaxies.
A decision tree is built up through a recursive training process as follows. 
First, the bootstrap sample is divided into two sub-samples,
left child $\mathcal{D}_{\rm bts, L}$ and right child $\mathcal{D}_{\rm bts, R}$,
according to a critical value of one feature. The feature and the 
critical value are both chosen to minimize the Gini impurity, 
which is defined as
\begin{equation}
    {\rm Gini} = \sum_{k={\rm L, R}}\frac{|\mathcal{D}_{{\rm bts}, k}|}{|\mathcal{D}_{\rm bts}|}
	\left(1 - \sum_{i = 1}^{|\mathcal{Y}|}p_{k,i}^2\right)
\end{equation}
where $|\mathcal{D}_{\rm bts}|$ is the size of the bootstrap sample, 
$|\mathcal{D}_{\rm bts, k}|$ is the size of the sub-sample, 
$|\mathcal{Y}|$ is the dimension 
of the target space (number of target classes), and 
$p_{k,i}$ is the fraction of the $i$-th class objects in the sub-sample $k$. 
A small value of the Gini impurity, therefore, indicates  
high purity of the target vectors in each of the sub-samples. 
This process is repeated for each sub-sample recursively until some 
termination criterion is met. Each splitting is 
referred to as an internal node, and a sub-sample that will 
not be split further is called a leaf node. A termination criterion 
can be set to achieve either a user-defined maximum depth of the tree,  
or a minimal sample size (number of photometric galaxies) required for further splitting.
Each leaf node is assigned a target vector specified as the mode of 
the target vectors of the objects it contains. After the training process, 
each decision tree can be used to predict the target vector for any other 
input object by assigning it to a leaf node according to its feature values. 
Finally, since each of the decision trees (i.e. each of the bootstrap samples)
gives a target vector prediction for an input object, the RFC chooses 
the mode of the target vectors as the final prediction for the object.
	
Several hyper-parameters are used to control the flexibility of the RFC: 
\texttt{n\_estimators} specifies the number of decision trees; 
\texttt{min\_samples\_split} specifies the minimal number of objects
for further splitting an internal node; \texttt{max\_features} specifies
the number of features chosen for each bootstrap sample;
and \texttt{class\_weight} specifies the weight of training samples with 
different target values.

We create 20 different mock samples for our tests (see below). 
For each mock, we combine photometric galaxies from other five mock samples to form a
training sample and apply the trained RFC to that mock
(recipient sample). This process is repeated in turn for 20 different combinations 
of training and recipient samples, so that we have RFC predictions 
for all the 20 mock samples to test the accuracy of the classification. 

\section{Mock catalogs}%
\label{sec:mock_catalogs}

\subsection{Source selection}%
\label{sub:source_selection}

\begin{table}
	\centering
	\caption{The PFS survey selection criteria.}
	\label{selection}
	\begin{tabular}{lccr}
		\hline
		Redshift      &$m_{\rm limit}$       & Sampling rate\\
		\hline
		0.7 < z < 1.0 & y < 22.5             & 50\%\\
		1.0 < z < 1.7 & y < 22.5             & 70\%\\
		1.0 < z < 1.7 & y > 22.5 \& J < 22.8 & 70\%\\
		\hline
	\end{tabular}
\end{table}
	
To quantify the performance of the group finder described above,
we have constructed mock catalogs which mimick existing and future
high-$z$ galaxy redshift surveys. Detailed description of the
mock catalogs can be found in a parallel paper by \citet{mengMeasuringGalaxyAbundance2020}.
These catalogs are based on ELUCID
\citep{wangELUCIDEXPLORINGLOCAL2016}, a large $N$-body cosmological
simulation  run with $3072^3$ particles in a box of $500{\rm Mpc/h}$ on a side.
Dark matter halos are populated with galaxies using an
empirical model of galaxy formation, constrained by the local stellar
mass function of galaxies in rich  clusters and the stellar mass
function of galaxies from $z=0$ to $5$  \citep[see][for
details]{luEmpiricalModelStar2014}. The implementation of the
empirical model in the simulated ELUCID  halo merger trees is
described in \citet{chenELUCIDVICosmic2019}. 
The minimal halo mass is about $\rm 10^{10}M_{\odot}/h$ 
in the simulation, but the merger trees are extended
to $\rm 10^9M_{\odot}/h$ using a Monte Carlo method. 
The corresponding minimal stellar mass is about 
$\rm 10^8M_{\odot}/h$, much lower than the PFS and 
zCOSMOS-bright targets.
Light-cone  mock catalogs are constructed by \citet{mengMeasuringGalaxyAbundance2020}
to mimic the selection criteria of galaxy redshift surveys at 
intermediate and high redshifts, such as zCOSMOS-bright \citep{Knobel:2012je} and 
the upcoming Prime Focus Spectrograph (PFS) galaxy survey 
on Subaru \citep{takadaExtragalacticScienceCosmology2014}.

The PFS survey will be carried out by the 8-meter Subaru telescope, with 
the spectroscopy to be obtained with 2,394 fibers distributed in a hexagonal 
field of view with an effective diameter of about 1.3 degree. 
As one of three major experiments of the PFS project, the PFS 
galaxy evolution survey will obtain spectroscopy for about 256,000 galaxies 
over the redshift range from $z=0.7$ to 1.7 and a sky coverage of 
$\sim 14.5\ {\rm deg}^2$ (see Table\,\ref{selection} for the 
PFS galaxy target selection criteria).
The redshift sampling rate ranges from 
50\% to 70\% in different redshift ranges, so that about 30 -- 50\% of the 
galaxies that meet the target selection criteria will not have
spectroscopic observation. This will affect the 
completeness of the group catalog to be constructed, as we will see below. 
To reduce the impact of such incompleteness,  
we will use photometric data from the Hyper Suprime-Cam SSP survey,
which is complete to $\rm y=25.3$ \citep[][]{aiharaHyperSuprimeCamSSP2018}.
For galaxies satisfying the selection criteria in Table \ref{selection}
and having no spectroscopic redshift measurements, we will use their photometric
redshifts, which have an accuracy of $\Delta z/(1+z)\sim 0.02$.
To quantify cosmic variances, we generate 20 different mock samples 
from the simulation. These mock samples are constructed with random tiling 
and shifting of the simulation box so as to minimize duplicates of 
structures among them. In \citet{mengMeasuringGalaxyAbundance2020},
it is shown that the covariances in the number density and 
clustering of galaxies between different mock samples are 
much smaller than the variances, indicating that 
these mocks may be considered as independent statistically.

\subsection{Sampling effect}%
\label{sub:sampling_effect}
	
Due to the limited number of fibers on the focal plane, one has to revisit the 
same pointing several times in order to achieve the planned sampling rate. 
For the PFS project, the sampling effect can be mimicked 
using the fiber assignment software, 
Exposure Targeting Software 
(ETS)\footnote{\href{https://github.com/Subaru-PFS/ets_fiber_assigner}{https://github.com/Subaru-PFS/ets\_fiber\_assigner}.}, 
which is being developed by the PFS collaboration. In our modeling, we tune the number of 
visits for each pointing to ensure the average sampling rate 
listed in Table\,\ref{selection}. Since most of the survey volume is enclosed 
by the redshift range from 1.0 to 1.7, we only consider galaxies in this redshift 
range when testing our group finder.  
The corresponding sample produced by the ETS will be denoted as ETS($f$), 
and we only consider $f=70\%$ as an example. 
	
Although the mock catalogs described above are created 
for the PFS galaxy evolution survey, we will use the parent sample to construct 
a set of more general mock catalogs that may be applicable to 
other deep redshift surveys, such as zCOSMOS \citep{Lilly:2009fb}, 
DEEP2 \citep{Newman:2013ha}, and VVDS \citep{LeFevre:2005bs}.
As mentioned earlier, limited spectroscopic sampling 
is a common property of these deep redshift surveys. 
To quantify the effects of such incompleteness on group identification, 
we construct mock catalogs with a set of different sampling rates 
denoted as Rand($f$) where $f=100\%, 85\%, 70\%, 55\%$, respectively.
The catalog of a given sampling rate 
is obtained  by randomly selecting the corresponding  
fraction of galaxies from the complete parent sample.

In general, the final sampling effect is determined by the combination 
of two types of sampling processes. First, the spatial sampling process, 
e.g. fiber assignment, determines which galaxies are targeted 
by the spectral observation among all the sources that satisfy 
the selection criteria. This effect is spatially inhomogeneous 
and may depend on the distribution of galaxies in the sky. 
The other effect is called redshift success rate, i.e. the probability 
to accurately determine the redshifts from the observed spectra. 
The latter effect may depend on the luminosity, redshift or color of 
the sources. In both cases, the incompleteness can be described 
by an incompleteness map which specifies the probability for
the target objects to be included in the spectroscopic sample. 
As demonstrated in \citet{mengMeasuringGalaxyAbundance2020}, 
our mock catalogs not only reproduce the general population of galaxies in the 
redshift range probed in terms of both abundance and clustering, 
when compared to the real galaxy samples provided by the zCOSMOS survey,  
but also mimic the selection effects that are generally applied 
to real surveys at high redshift.  Therefore, these mock catalogs 
can be used here for the purpose of testing our group finding algorithms. 

\section{Testing the performance of the group finder}%
\label{sec:testing_the_performances_of_the_group_finder}
	
\subsection{Performance measures}%
\label{sub:performance_measures}

A good group finder should correctly identify a high fraction of true 
groups, and simultaneously include a low fraction of false groups 
which are not true groups.
We define two quantities to characterize the performance of our group finder:
{\em completeness} and {\em purity}.
Completeness is defined as the fraction of true groups that are 
correctly identified by the group finder, and  purity
is defined as the fraction of all the identified groups that are true. 
For convenience, we use the following two terms in our 
description: 
{\bf Identified Group (IG)}, defined as a group identified by the group finder; 
{\bf True Group (TG)}, defined as a true group in the mock catalog. 
In practice, it is  not straightforward to match IGs 
with the corresponding TGs. This is because in many cases 
an IG is composed of a portion of the member galaxies of the 
corresponding TG plus a number of interlopers, while the member
galaxies of a TG may be divided into different IGs.
Here we consider three matching schemes that we will use to
link IGs and TGs:
\begin{enumerate}
\item {\it Member Matching (MM)}.
	The MM scheme was called {\it two-way matching} in \citet{knobel2009optical}.
    This matching is established if more than $\phi\times N_I$ members in an IG 
	belong to the same TG, and more than $\phi\times N_T$ members in this TG is 
	contained by the IG. Here $N_T$ is the richness of the TG modified by the 
	sampling process, and $N_I$ is the richness for the IG.
        For $\phi\ge 0.5$, this scheme leads to a perfectly one-to-one matching,
        and we thus adopt $\phi=0.5$. However, 
        this matching scheme may too strict for poor systems, 
        where incorrect assignments of a few low-mass members may not 
        affect much the halo mass calibration, but can   
        change $N_I$ significantly so as to affect the match between 
        IG and TG.
\item{\it Central Matching (CM)}.
	The matching is established if the central galaxy of a TG is correctly 
	identified as the central of an IG. This matching criterion is used by 
	\citet{lim2017galaxy}. Because of incomplete sampling, an IG
	can have its central lost while still keeping many of its satellites 
	in the spectroscopic sample. Such systems cannot be matched in 
	the CM scheme.
\item{\it Member or Central Matching (MCM)}. In this case, we
  combine the MM and CM schemes to overcome the problems of the previous
  two matching schemes, and we refer this new scheme as {\it Member or Central 
    matching}. The matching is established if a TG and an IG satisfy 
	either the MM or the CM scheme. If a TG (or an IG) is matched with 
	two counterparts, the MM pair has the priority. 
	This matching scheme is one-to-one, 
	as the previous two matching schemes. We will adopt this matching 
	scheme in what follows.
\end{enumerate}

If an IG is matched with a TG, the IG is said to be true, and is referred 
to as an IG-T. Similarly, if a TG is matched with an IG, the TG is said to be 
identified, and is referred to as a TG-I.

With the matching scheme above, we define the completeness in two ways. 
The first one, $C_1(N)$, introduced in \citet{knobel2009optical}, 
is defined as the fraction of TG-Is 
among all TGs in the mock catalog (including the effect of incomplete sampling) 
as a function $N$, where $N$ is the richness of a galaxy group 
obtained from the incomplete sample. The maximum value of 
$C_1$ is 1.0. The second, $C_2(M_h)$, 
is defined as the fraction of TG-Is of given mass, $M_h$, among all halos 
of such mass in the volume of the mock catalog (without including
the incomplete sampling). The maximum value of $C_2$ is limited by the sampling 
rate, as we will see later. For the purity, $P$, we also use the definition of 
\citet{knobel2009optical}, which is the fraction of IG-Ts among all the IGs 
in the group catalog as a function of richness. 
Table\,\ref{symbols} lists the acronyms and quantities defined above.
	
\begin{table}
	\caption{Summary of symbols used in this paper} 
	\label{symbols}
	\centering
	\begin{tabular}{c|c|}
		\hline    
		Symbol           & Interpretation\\
		\hline   
		TG               & True galaxy group in the mock\\
		\hline
		IG               & Identified galaxy group with the group finder \\
		\hline 
		TG-I             & \tabincell{c}{True galaxy group which is matched with \\ identified galaxy group under MCM matching scheme} \\
		\hline   
		IG-T             & \tabincell{c}{Identified galaxy group which is matched with \\true galaxy group under MCM matching scheme} \\
		\hline
		$C_1(N)$         & $\frac{\text{\# of TG-Is}}{\text{\# of TGs in the sampled mock}}$ as function of richness $N$\\
		\hline
		$C_2({\rm M_h})$ & $\frac{\text{\# of TG-Is}}{\text{\# of halos in the survey volume}}$ as function of halo mass $M_h$\\
		\hline
		$P(N)$           & $\frac{\text{\# of IG-Ts}}{\text{\# of IGs in the sampled mock}}$ as function of richness $N$\\
		\hline
		Rand($f$)        & \tabincell{c}{Mock with sampling rate as $f$\\ by proceeding the sampling process randomly}\\
		\hline
		ETS($f$)         & \tabincell{c}{Mock with sampling rate as $f$\\ by proceeding the sampling process using ETS software}\\
		\hline
	\end{tabular}
\end{table}

\subsection{Performance on spectroscopic samples}%
\label{sub:performance_on_spectroscopic_samples}

As described in \S\ref{sub:the_friend_of_friend_method},
the FoF group finder contains three free parameters that need 
to be calibrated: $l_{\mbox{max}}$, $b$ and $R$.
Motivated by the quantity 
$\tilde g_1$ defined in \citet{Knobel:2012je}, 
we calibrate these parameters by minimizing the following quantities,
\begin{equation}
	g = \sqrt{\left[1 - \bar C_1(1-10)\right]^2 
	        + \left[1 - \bar P(1-10)\right]^2}
\end{equation}
where $\bar C_1(1-10)$ and $\bar P(1-10)$ represent the average values of 
$C_1$ and $P$ for systems with richness from $N=1$ to $10$ under 
the MCM scheme. We find that the optimal parameters for different 
sampling cases are quite similar. For simplicity we therefore 
use the same set of parameters, as given in Table~\ref{para_finder},  
for all the sampling cases. We note that the difference in the results 
obtained from the optimal parameter set and the set adopted is small. 
	
\begin{table}
	\centering
    \caption{Adopted FoF parameters calibrated with mock PFS samples.}
	\label{para_finder}
	\begin{tabular}{c|c|c|c}
		\hline
		Parameters & $b$  & $l\_{\rm max}$(Mpc/h) & $R$   \\
		\hline\hline
		Values     & 0.09 & 0.25                  & 19.13 \\
		\hline
	\end{tabular}
\end{table}
	
With the three parameters determined, we apply the group finder
to 20 different mock catalogs. The performances on the group level 
are shown in Fig.\,\ref{group_performance_pfs} as dashed lines. 
As one can see, for cases of random sampling, both the $C_1$ and $P$ 
indices can reach 90\% even for a sampling rate as low as 55\%. 
This indicates that the FoF method can identify 
most of the galaxy systems and that the identified groups are mostly true. 
Meanwhile, the $C_2$ index decreases systematically with decreasing 
sampling rate. The decrease is larger for systems of lower masses.
The reason for this is simple: halos of lower masses typically 
contain smaller number of member galaxies, so that the probability 
for them to lose all their members in the spectroscopic sample is higher. 
For ETS(70\%), both $C_1$ and $P$ can still reach 90\%, but the $C_2$ 
index is lower than that in Rand(70\%), especially for rich/massive 
groups. This happens because the ETS fiber assignment algorithm makes
the sampling rate lower in higher density regions where rich/massive
systems are usually located.
	
\subsection{Improvement by incorporating photometric data}%
\label{sub:improvement_by_incorporating_photometric_data}

\begin{table}
	\centering
	\caption{Adopted hyper-parameters of the random forest classifier for photometric galaxy classification.} 
	\label{hyperparameter}
	\begin{tabular}{|c|c} 
		\hline       
		Hyper-parameter              & Value\\
		\hline\hline
		\texttt{n\_estimators}       & 30 \\
		\texttt{min\_samples\_split} & 10 \\
		\texttt{max\_features}       & 6 \\
		\texttt{class\_weight}       & \texttt{balanced}\\
		\hline
	\end{tabular}
\end{table}

\begin{figure*}
	\includegraphics[width=\linewidth]{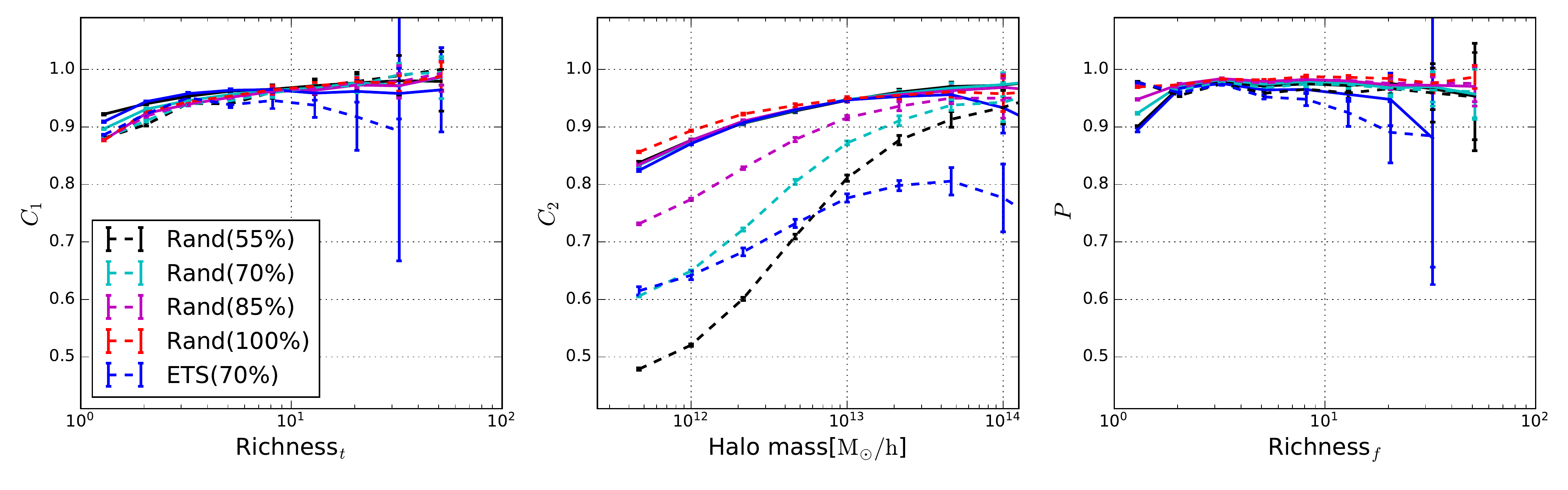}
	\caption{Performance comparison under MCM scheme with and without 
	photometric data. {\it Solid lines} are results with photometric data
	while {\it dashed lines} use only spectroscopic data. Error bars are 
	the standard deviations among 20 different mocks.}
	\label{group_performance_pfs}
\end{figure*}

The good performance of the FoF group finder in terms of $C_1$ and $P$
indicates that the group finder is able to correctly identify most of the 
galaxy systems that are contained in the spectroscopic sample. 
Thus, the low $C_2$ values for cases of low sampling rates 
must be due to the missing of group systems in the  
spectroscopic sample, caused by incomplete sampling of the survey.
In order to find these lost systems, we make use of 
information from the parent photometric sample. As mentioned in \S\,\ref{sec:method}, 
we apply the RFC to identify two kinds of lost
central galaxies from the photometric sample:
the {\it group central} and the {\it isolated central}.

To determine if a photometric galaxy is a {\it group central},
or an {\it isolated central}, or neither of the two, we 
characterize the relationship between 
the photometric galaxy and the spectroscopic groups around it. As described in 
\S\,\ref{sub:supplementing_with_photometric_galaxies}, we do this by determining both
the hyper-parameters for RFC and $n_g$, the number
of spectroscopic groups around the galaxy in question.

To find the optimal hyper-parameters of the RFC, we employ the $n$-fold 
cross-validation method. First, we randomly divide the photometric sample into $n$ 
sub-samples with an equal number of galaxies. We then train the model
on $n-1$ sub-samples 
and make a prediction for the remaining one to test the performance. 
This process is repeated for each of the $n$ sub-samples. Here we choose $n=5$.
We use the following set of quantities to describe the goodness 
of the prediction: 
\begin{itemize}
\item $C_{\rm iso}$:
    Completeness of isolated centrals, 
    defined as the fraction of isolated centrals that are 
    correctly identified among all the isolated centrals in the photometric sample;
\item $P_{\rm iso}$:
    Purity of isolated centrals, 
    defined as the fraction of isolated centrals which are correctly identified 
    among all the found isolated centrals;
\item $C_{\rm grp}$:
    Completeness of group centrals, 
    defined as the fraction of group centrals correctly identified among all group centrals, 
    where a group central is the central of a group with 
    at least one galaxy in the spectroscopic sample;
\item $P_{\rm grp}$:
    Purity of group centrals, 
    defined as the fraction of group centrals correctly identified among all the 
    identified group centrals. 
\end{itemize}
The hyper-parameters are chosen to achieve a balance among the 
above four quantities. Specifically, we optimize the values of the 
hyper-parameters by maximizing the quantity, $g$, defined as 
\begin{equation}
	g = C_{\rm iso}\cdot P_{\rm iso}\cdot C_{\rm grp}\cdot P_{\rm grp}
\end{equation}
We find that the $g$ index is not sensitive to the exact 
values of the hyper parameters, so we will use the set of hyper-parameters
given in Table\,\ref{hyperparameter} for different cases of redshift sampling.
We also find that $n_g=3$ is sufficient for our purpose, independent of the 
redshift sampling. 

Using the updated group catalog that incorporates photometric data, 
we plot the performance of the MCM scheme in 
Fig.\,\ref{group_performance_pfs}. It can be seen that the main 
improvement is in the $C_2$ index at the low-mass end. 
This happens because most of the isolated centrals that are lost 
in the spectroscopic sample are now found in the 
photometric data. In addition, the missed massive groups 
in the ETS(70\%) case can also be identified from the photometric data. 
There is, however, a noticeable decline in the purity at 
${\rm Richness}_f = 1$, since not all the isolated 
centrals identified from the photometric data are true centrals.

\subsection{Assigning Halo Masses to Groups}%
\label{sub:assigning_halo_masses_to_groups}

\begin{figure*}
	\includegraphics[width=\linewidth]{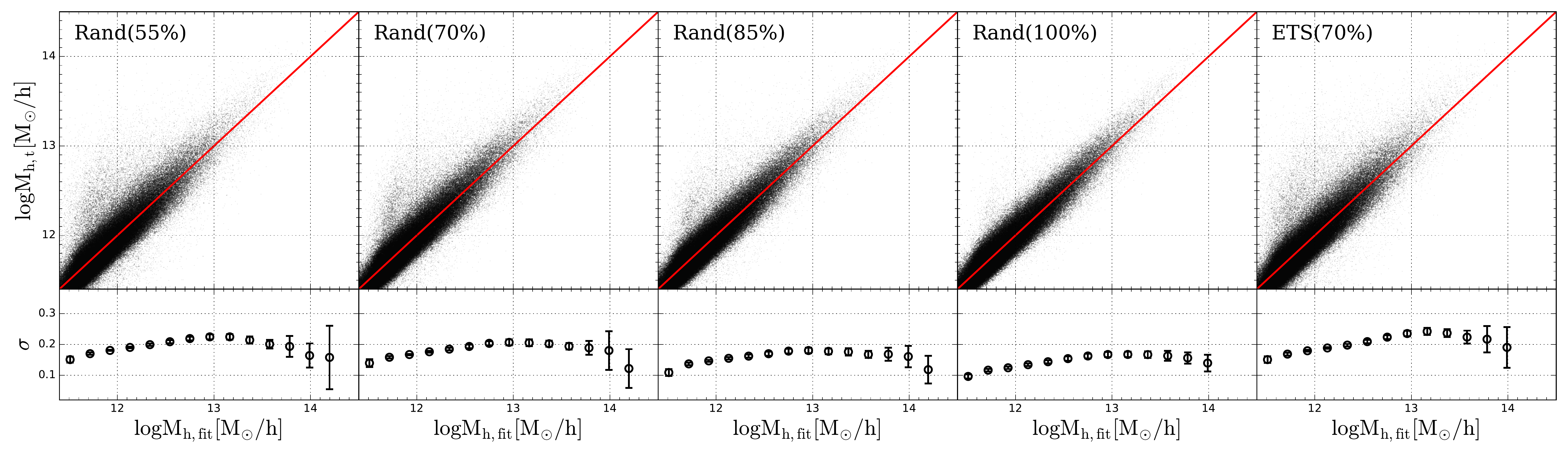}
	\caption{Upper panels: the scatter plot between the predicted halo 
		mass and the true halo mass. Lower panel: the standard deviation 
		in the true halo mass for a given predicted halo mass, with error 
		bar showing the variance among 20 mocks.}
		\label{halomass_performance_pfs}
\end{figure*}

\begin{figure*}
	\includegraphics[width=\linewidth]{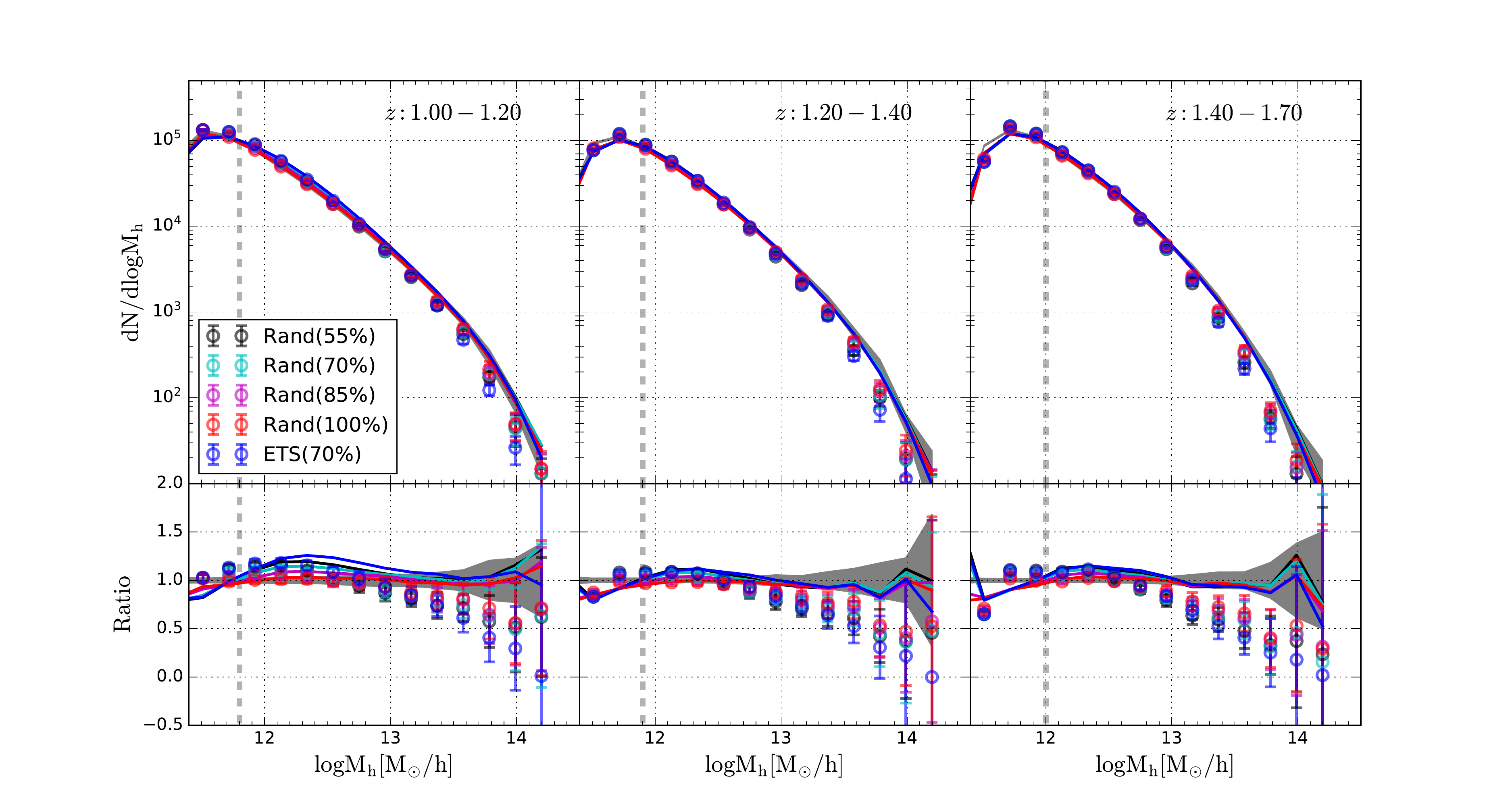}
	\caption{Halo mass functions in three redshift bins. 
		The circle in upper panels are the halo mass functions of groups identified 
		from samples of different redshift samplings, with error bars representing
        the variance among 20 mocks. The solid lines are means of the 
        distribution for $\rm M_{h, samp}$.
        The shaded areas are the mass functions of simulated halos used 
        to construct the mock catalogs, with the width indicating the variance 
        among 20 mocks. The vertical dashed lines indicate
        masses below which the halo samples become incomplete. 
        The lower panel shows the ratio of halo mass distribution obtained from 
        the identified groups to that of the simulated halos.}
	\label{halomassfunction_pfs}
\end{figure*}

Galaxies are formed and evolved in dark matter halos, and so 
the total stellar mass and number of member galaxies in a host halo 
are expected to be related to the dark matter mass of the host halo.
Thus, it is possible to infer the halo mass of a group 
from the galaxies it contains. In this subsection, we apply
the Random Forest Regressor (RFR), which is similar to the RFC, 
to infer the host halo mass for each of the identified galaxy
groups \citep[see][for a recent application of the RFR in this regard]{manFundamentalRelationsHalo2019}. 
The RFR is different from RFC in two ways.
First, instead of the Gini impurity, RFR partitions the feature 
space to minimize the mean squared error (MSE), defined as 
\begin{equation}
    {\rm MSE} = \sum_{j=1}^{|\mathcal{D}_{\rm bts, L}|}\left(y_j - \bar y_{\rm L}\right)+\sum_{j=1}^{|\mathcal{D}_{\rm bts, R}|}\left(y_j - \bar y_{\rm R}\right)
\end{equation}
where $|\mathcal{D}_{\rm bts, L}|$ and $|\mathcal{D}_{\rm bts, R}|$
are the sizes of the two sub-samples at a node, 
$y_j$ is the $j$-th target value, and $\bar y_{\rm L}$ 
and $\bar y_{\rm R}$ are the means of the target values
in the tow sub-samples.
Second, the target value for each leaf is chosen 
to be the mean target value of the training sample in each 
leaf, rather than the mode. We use the following features 
from both the spectroscopic and photometric data to infer the halo mass:
\begin{enumerate}
\item ${\rm M_{*,tot}}$:
	the total stellar mass;
\item ${\rm M_{*,c}}$:
	the stellar mass of the central galaxy;
\item ${\rm N_{\rm tot}}$:
    the group richness, which is the total number of
    member galaxies (both spectroscopic and photometric);
\item ${\rm \sigma_G}$:
    velocity dispersion estimated using the gapper
    algorithm \citep[][]{beersMeasuresLocationScale1990},
    \begin{equation}
        \sigma_G = {\sqrt{\pi} \over N(N - 1)}\sum_{i = 1}^{N-1}i(N-i)(v_{i + 1} - v_i)
    \end{equation}
    where $v_i=cz_i/(1 + z_{\rm grp})$, with  
    $v_1 \le v_2 \le ... \le v_N$, are the velocities of 
    spectroscopic members, $z_{\rm grp}$ is the mean of $z_i$,
    and $N$ is the number of spectroscopic members.
    We set $\sigma_G=-1$ for systems with $N<2$.
\item ${\rm group~tag}$:
	which is equal to $0$ for a pure 
    spectroscopic group, $1$ for a group with photometric central 
	and spectroscopic members, and $2$ for an isolated photometric central;
\item ${\rm Redshift}$:
    group redshift, defined to be the photometric redshift 
    of the central for groups that contain only a single 
    photometric central, and to be the 
    mean redshift of spectroscopic members for other groups;
\item ${\rm\log [M_{*, enc}(<5Mpc/h) - M_{*,tot}]}$:
    where $\rm M_{*,enc}(<5Mpc/h)$ is the total stellar mass of galaxies 
    whose projected distance to the group center (defined by the 
    sky position of the central and the redshift of the group)
    is smaller than 5 Mpc/h and the redshift difference 
    (using spectral $z$ and photo-$z$ for spectroscopic and 
    photometric galaxies, respectively) is smaller 
    than 3$\sigma_{z, \rm phot}(1 + z)$;
\item ${\rm\log [M_{*, enc}(<10Mpc/h) - M_{*,tot}]}$:
    similar to quantity defined above, except the projected distance 
    to the group is smaller than 10Mpc/h.
\end{enumerate}
As shown in the appendix, the information about halo mass is dominated 
by the first four features. 

The hyper-parameters are tuned to minimize the mean squared error of the 
halo mass. Here we employ the $n$-fold cross-validation method as in 
\S\,\ref{sub:improvement_by_incorporating_photometric_data}. 
The optimal values of the 
hyper-parameters are almost the same for different cases. We thus 
use the same set of values as given in Table\,\ref{hyperparameter_hm}
for cases of different redshift samplings. 

\begin{table}
	\centering
	\caption{Adopted hyper-parameters for Random Forest Regressor in halo mass calibration}
	\label{hyperparameter_hm}
	\begin{tabular}{|c|c} 
		\hline       
		Hyper-parameter              & Value\\
		\hline\hline
		\texttt{n\_estimators}       & 30 \\
		\texttt{min\_samples\_split} & 30 \\
        \texttt{max\_features}       & 3 \\
		\hline
	\end{tabular}
\end{table}

\begin{figure*}
	\includegraphics[width=\linewidth]{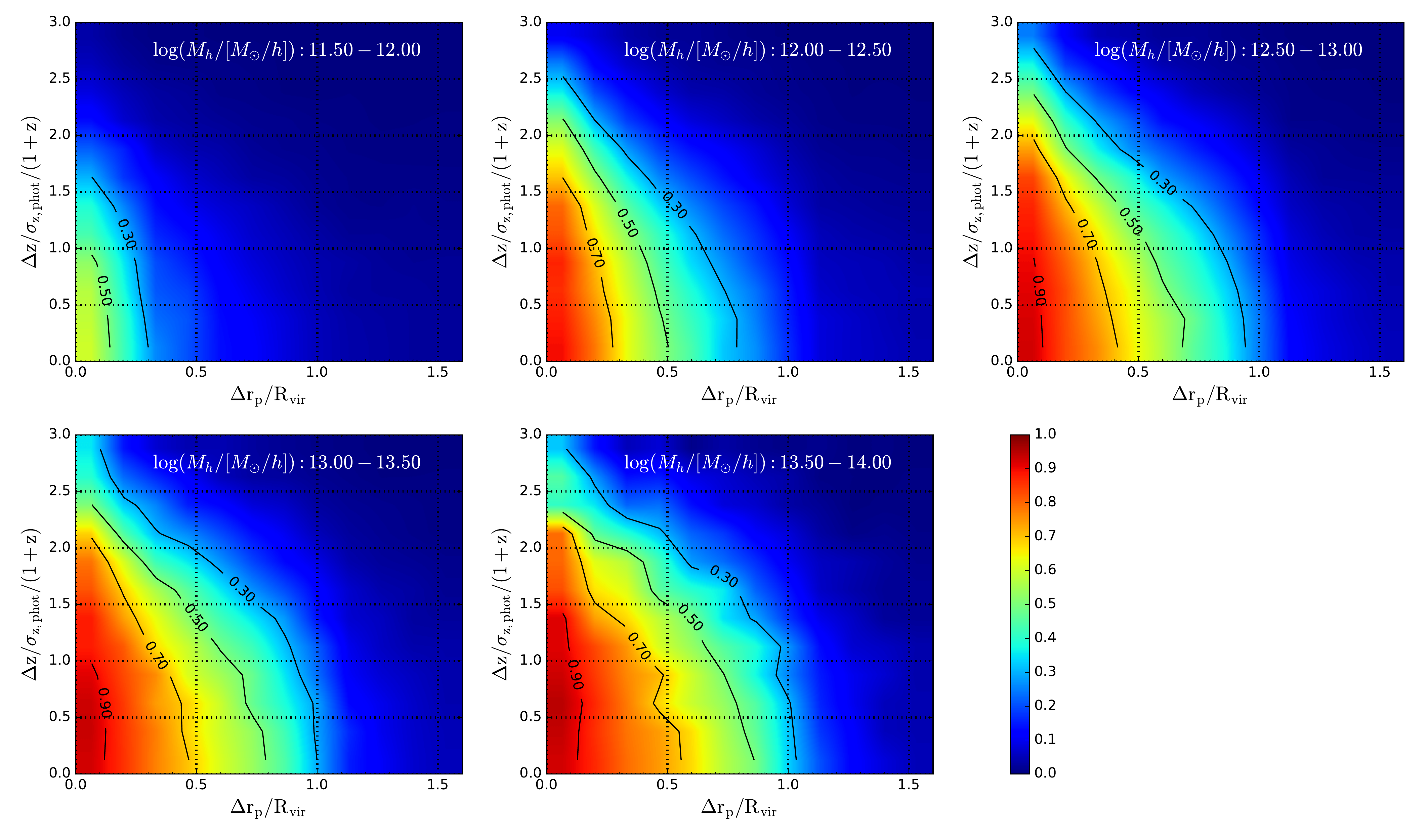}
	\caption{Fraction of true members as a function of 
	$\Delta r_p$ and $\Delta z$ to the central galaxy. The figure is for Rand(55\%).}
	\label{frac_map}
\end{figure*}

\begin{figure*}
	\includegraphics[width=\linewidth]{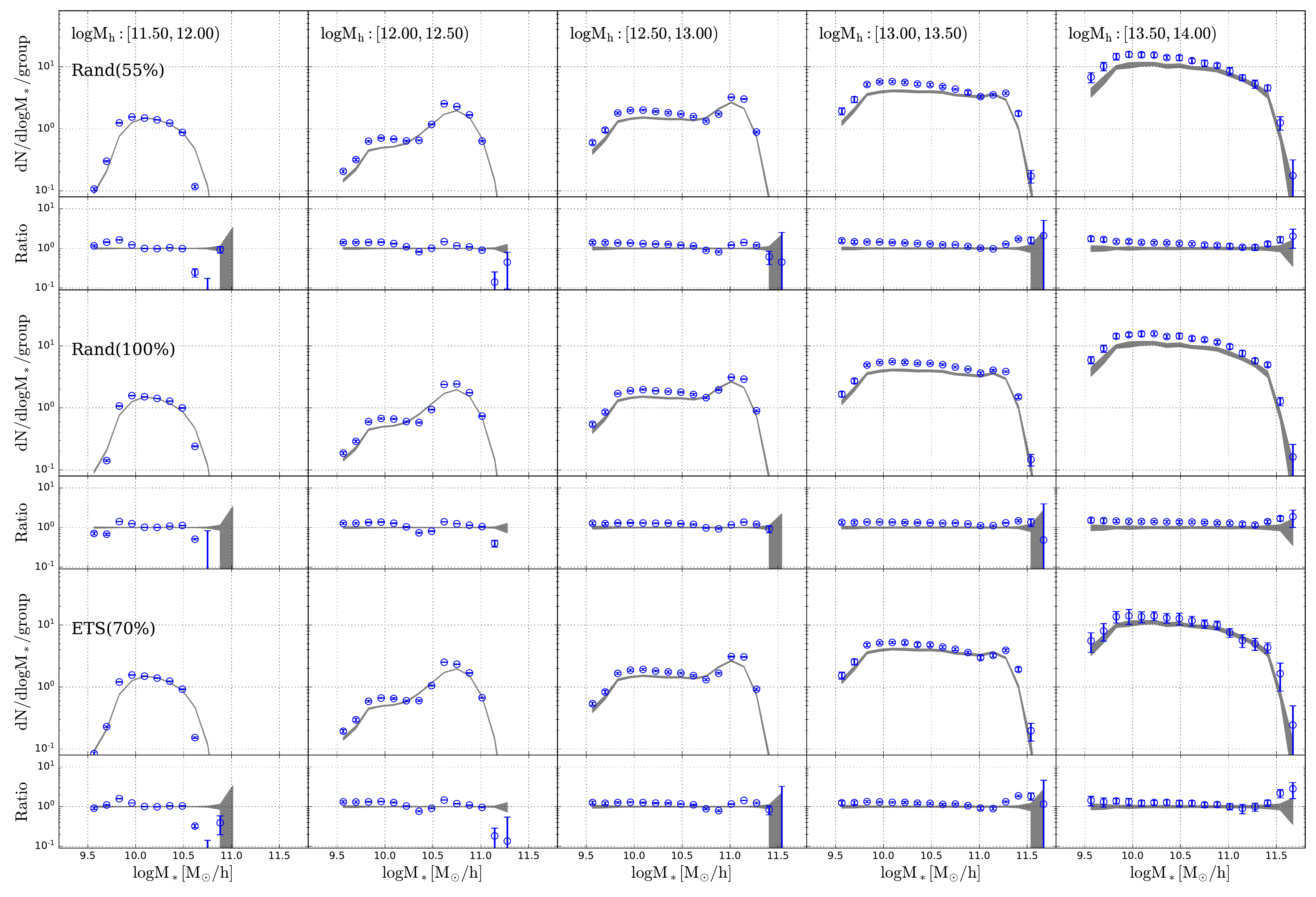}
	\caption{Conditional stellar mass functions in five halo mass bins
		obtained from samples of different redshift sampling rates.
		Blue circles are obtained from identified groups
        with error bars representing variation in 20 mocks (see text). 
        And gray regions are obtained from model galaxies in simulated halos,
        with width represents the variance among 20 mocks.
        We also plot the ratio of the measurements to the mean value
        of the CSMF of model galaxies in simulated halos in the small panels.}
	\label{csmf}
\end{figure*}

\begin{figure*}
	\includegraphics[width=\linewidth]{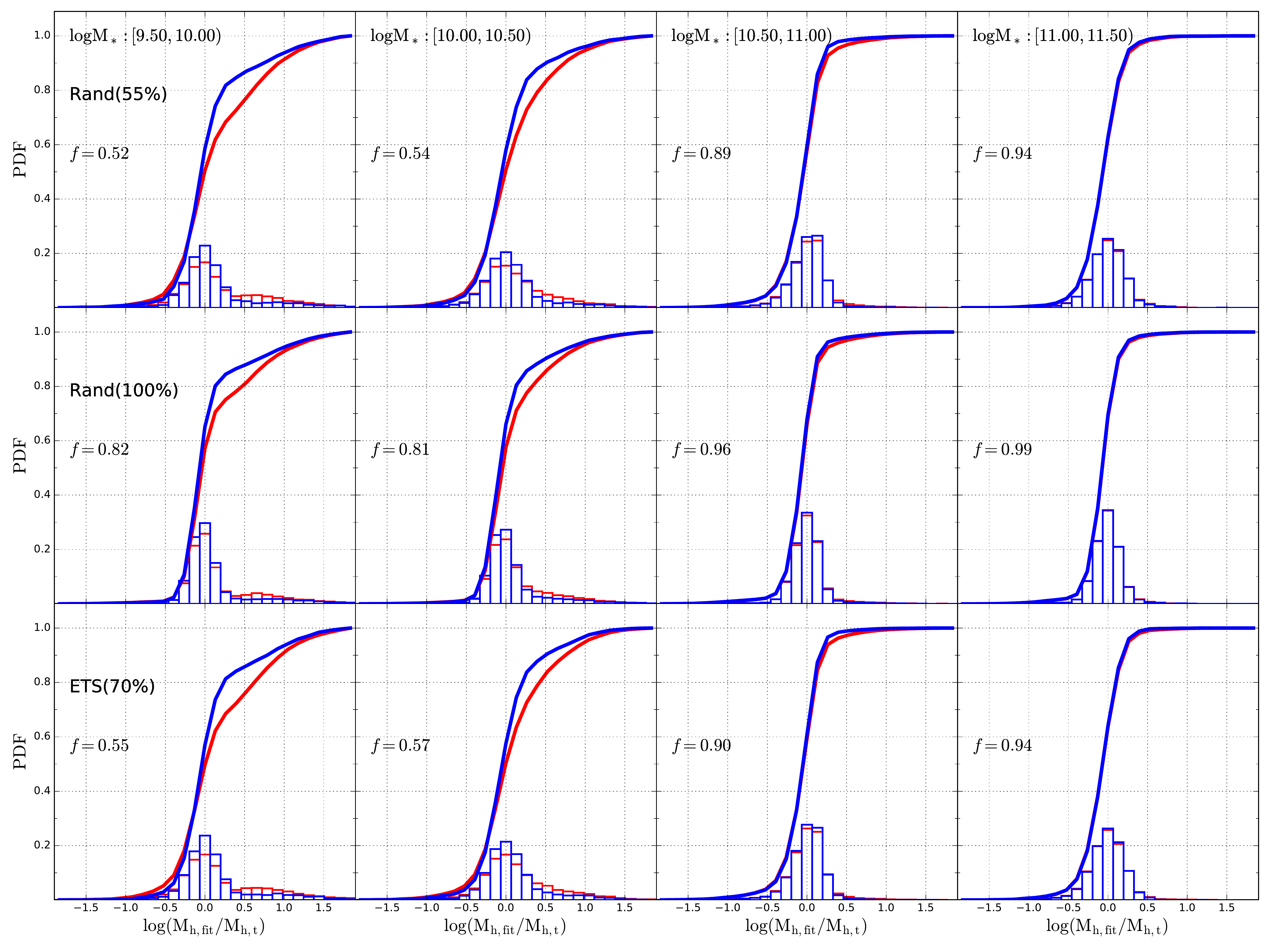}
    \caption{Blue histograms: distribution of $\rm\log(M_{h, fit}/ M_{h, t})$ for 
    galaxies with $\rm M_{h, fit} > 10^{12}M_{\odot}/h$; Red histograms: 
    distribution of $\rm\log (M_{h, fit} / M_{h, t})$ for galaxies with 
    $\rm\log (M_{h, fit}h/M_{\odot}) > 12$, and with
    $\rm \Delta r_p / R_{vir} < 1.0$ and $\Delta v / \rm V_{vir} < 2.0$ 
    (see text). The solid lines are the corresponding accumulated distribution. 
    The $f$ indicates the number ratio of galaxies in the blue
        histogram with that in the red.}
	\label{delta_hm_hist}
\end{figure*}

We use all the 20 mock catalogs to check the performance of the halo mass prediction. 
For each mock catalog, we use five other mock catalogs to train the RFR and 
to predict the results for the mock in question. The performance of the halo
mass prediction is quantified by the discrepancy between the true halo 
mass, $\rm M_{h, t}$, and the predicted (fitted) halo mass, $\rm M_{ h, fit}$.

In Fig.\,\ref{halomass_performance_pfs}
we plot the relation between $\rm M_{h, t}$ and $\rm M_{ h, fit}$
(upper panels) and the standard deviation of 
$\rm\log(M_{h, fit}/ M_{h, t})$ (lower panels),
for cases of different redshift sampling.
For the case of 100\% redshift 
sampling, the standard deviation ranges from 0.1 dex to 0.2 dex 
over the halo mass range from $\rm\sim 10^{11}M_{\odot}/h$ to 
$\rm \sim 10^{14}M_{\odot}/h$. This is similar to the result in
\citet{lim2017galaxy} for the SDSS galaxy sample using the halo-based 
group finder. For the cases of random sampling, the standard 
deviation at given halo mass increases with decreasing sampling rate,
reaching a range between 0.15 dex to 0.22 dex for the sampling rate of 55\%.
In the case of ETS(70\%), the overall performance is slightly worse than that of 
Rand(70\%), particularly at the massive end ($>10^{13}M_{\odot}/h$).
This can be understood as follows: due to fiber collisions
the effective sampling rate is a decreasing function of
galaxy target number density, leading to relatively low
sampling rates for massive systems which are located in high-density
regions. In ETS(70\%) the effective sampling rate is only about 30\% -- 40\%
at halo masses above $\sim 10^{13}M_{\odot}/h$.
As a result, many of the member galaxies in a massive group only
have photometric redshifts and cannot be assigned to the group reliably.
In addition, for cases with low sampling rates and for ETS(70\%), 
there are outliers at the low-mass end, caused by groups that can 
be identified but their halo masses are poorly predicted  
owing to the missing of member galaxies in the spectroscopic sample. 
	
The distribution of the predicted halo mass, $\rm M_{h, fit}$, is 
presented in Fig.\,\ref{halomassfunction_pfs} for three successive
redshift intervals over $1<z<1.7$, in comparison with
the halo mass functions obtained directly from the simulation 
used to construct the mock catalogs.
It is obvious that the halo mass distribution is under-estimated to
varying degree at the massive end ($>10^{13}M_{\odot}/h$),
even for Rand(100\%). This is expected, 
because our halo mass estimate is optimized for 
each selected group to have an estimated mass ($\rm M_{h, fit}$) 
that best match the true mass ($\rm M_{h, t}$), 
and because there is scatter 
in the true halo mass for a given estimated mass
(see Fig.\,\ref{halomass_performance_pfs}). 
To take into account the effects of such scatter, 
we introduce a random variable, $\rm M_{h, samp}$, defined as
\begin{equation}
    \rm M_{\rm h, samp} = M_{\rm h, fit} + {\rm Norm}[0, ~\sigma(M_{\rm h, fit})]
    \label{hm_samp}
\end{equation}
where $\rm Norm[0,~\sigma(M_{h,fit})]$ is a random number generated 
from a normal distribution with zero mean and a standard deviation, 
$\rm \sigma(M_{h, fit})$, as inferred from Fig.\,\ref{halomass_performance_pfs}. 
To estimate a statistical quantity, $s$, using a set of
halo masses, $\{{\rm M_{h, fit}}\}$, we first generate a set of halo 
masses, denoted by $\{{\rm M_{h, samp}}\}_i$, 
using Eq.(\ref{hm_samp}), and repeat the process $N_{\rm samp}$
times. Our estimate for $s$ is
\begin{equation}
    s_{\rm samp} = {1\over N_{\rm samp}}\sum_{i = 1}^{N_{\rm samp}}s(\{{\rm M_{h, samp}}\}_i).
\end{equation}
The average distribution of $M_{\rm h, samp}$, obtained using $N_{\rm samp} = 30$,
is calculated in this way and plotted in Fig.\,\ref{halomassfunction_pfs}
as the corresponding solid line for each of the cases.
As one can see, the distribution of
$M_{\rm h, samp}$ matches well the true halo mass function in the simulation
for all cases, demonstrating again that the group sample 
selected by our group finder is quite complete and unbiased in 
the mass distribution. Note that due to the magnitude limit
in our mock galaxy sample, the halo sample selected is incomplete 
at the low-mass end. A halo mass limit, below which the 
incompleteness becomes significant is indicated by 
the vertical dashed line in Fig.\,\ref{halomassfunction_pfs}.
This limit is defined as the mass below which the amplitude of 
the estimated mass function deviates from the halo mass function
given by the original simulation by more than 0.05 dex.

\subsection{Group memberships}%
\label{sub:group_memberships}

The tests presented above are at the level of groups, based on  
group completeness and purity, and on halo mass assignments.
In this subsection we will test our group finder at the level of group 
members. We first consider the conditional stellar mass function 
(CSMF) of member galaxies in halos of a given mass, which is defined as the 
average number of member galaxies in these groups as a function of
the stellar mass of galaxies.
    
In order to account for redshift sampling effects, we need to include photometric 
galaxies around a group in a probabilistic way when calculating the CSMF. 
Here we employ a method similar to that proposed by \citet{knobel2009optical}, 
which consists of the following steps:
\begin{enumerate}
\item{\it Construct the map of the fraction of true members:}
	Using the mock catalog, we calculate the fraction of 
    true members among all the photometric galaxies, 
    excluding {\it group centrals} and {\it isolated centrals}, around 
    spectroscopic groups (those identified from spectroscopic galaxies 
    with spectroscopic or photometric centrals) 
    with given halo mass, in bins of the redshift difference, 
    $\Delta z/\sigma_{\rm z, phot}/(1 + z)$, and the projected separation, 
    $\Delta r_p /R_{\rm vir}$. As an example, Fig.\,\ref{frac_map} 
    shows the map of the fraction for the case of Rand(55\%).
\item{\it Assign membership probability:}
    After running the group finding pipeline, 
    each photometric galaxy, $i$, that has not been 
    identified as an {\it isolated central} or a {\it group central}, 
    will be assigned to a spectroscopic group, $J$, in its 
    neighborhood with a probability, $p_{i\rightarrow J}$,
    inferred from the fraction map constructed in previous step,
    based on the redshift difference and projected distance to the group.
    We note that each photometric galaxy, $i$, can be assigned 
    to several groups around in a probabilistic manner.
\item{\it Regulate the probability:} To ensure the summation of the 
    probabilities for a photometric galaxy to belong to all of its 
    neighboring spectroscopic groups and to be in the field
    is equal to one, we regulate the probability as \citep{Knobel:2012je}
	\begin{equation}
		\tilde{p}_{i\rightarrow J} = p_{i\rightarrow J}
            \times {{1 - p_{\rm field}}\over \sum_J p_{i\rightarrow J}},
            ~~~{\rm with}~~ p_{\rm field} = \prod_{J} (1 - p_{i\rightarrow J})
	\end{equation}
\end{enumerate}
Finally, we estimate the CSMF as
\begin{equation}
	\Phi (M_*\vert M_{\rm h, l},~M_{\rm h, u}) = {\sum_{i}\sum_{J} \tilde{p}_{i\rightarrow J}\over N_{\rm G}\Delta M_*}\,,
\end{equation}
where the summation on $i$ runs over all the galaxies whose stellar masses 
satisfy $M_{*} - \Delta M_*/2 \le M_{*,i} < M_{*} + \Delta M_*/2$, 
and summation on $J$ runs over all $N_{\rm G}$ spectroscopic groups whose 
halo masses satisfy $M_{\rm h, l} \le M_{*,j} < M_{\rm h, u}$.  
For each spectroscopic galaxy or {\it group central}, $i$, we set 
$\tilde{p}_{i\rightarrow J} = 1$ if it belongs to group $J$, and 
$\tilde{p}_{i\rightarrow J} = 0$ otherwise.

The CSMFs estimated in this way are plotted in Fig.\,\ref{csmf} 
in five halo mass bins (blue circles)
with error bars representing the variance between the 20 mock catalogs, 
in comparison with the CSMFs obtained 
directly from the member galaxies of dark halos in the simulation (gray shaded regions).
Here we only show results for three sampling cases since the
results of the other two cases fall in between Rand(55\%) and Rand(100\%).
As one can see, the CSMFs obtained from the identified galaxy groups
match well the input mock catalog. However, we overestimate slightly 
the amplitudes of the CSMFs at the low-mass end where the mass functions
are dominated by satellite galaxies.
This happens because we have adopted the same set of FoF parameters
calibrated with ETS(70\%),  which is slightly different from the 
optimal set for other cases of redshift sampling. The amplitudes 
of the CSMFs obtained from galaxy groups are also reduced if 
$M_{\rm h,samp}$ is used instead of $M_{\rm h,fit}$. 

Next, we consider the host halo mass distribution for spectroscopic galaxies
in four stellar mass bins. Different from the halo mass comparison for groups,
host halo mass distribution for galaxies are affected by membership 
assignment error, and thus provides a better quantification of halo 
mass uncertainties when halo masses are used as an environment indicator 
for individual galaxies. The differential and accumulated
distributions of $\rm\log(M_{h, fit} / M_{h, t})$ for all the 
spectroscopic galaxies in $\rm M_{h, fit} > 10^{12} M_{\odot}/h$ 
are presented in Fig.\,\ref{delta_hm_hist} as the red histograms and red 
solid lines, respectively. We note that there is a small tail in the 
distribution at high $\rm\log(M_{h, fit}/M_{h, t})$ for low stellar 
mass bins. This is produced by galaxies which are hosted 
by low-mass halos around massive groups but identified as 
satellites of the massive groups (interlopers) by the group finder.
To reduce the effects of these interlopers, one can trim the galaxy
sample by requiring the galaxies to satisfy the following criteria:
\begin{align}
    \Delta {\rm r_p} &< \alpha_r R_{\rm vir} \\
    \Delta v &< \alpha_v V_{\rm vir}
\end{align}
where $\Delta \rm r_p$ is the projected distance of a galaxy 
to the group center, and $\Delta v$ is the line of sight
velocity of the galaxy relative to the group center.
Here the group center is defined as the projected position of
central galaxy and mean redshift of spectroscopic members.
$R_{\rm vir}$ and $V_{\rm vir}$ are respectively the virial radius and virial 
velocity corresponding to the halo mass of the group $M_{\rm h, fit}$.
Blue histograms and blue solid lines in Fig.\,\ref{delta_hm_hist}
show the results for the case where $\alpha_r=1$ and $\alpha_v=2$. 
In each panel $f$ indicates the fraction of galaxies 
in the parent (untrimmed) sample that are kept after trimming. As expected,
the tail of the $\rm\log(M_{h, fit} / M_{h, t})$ distribution 
is largely reduced, especially at low stellar masses. Indeed, using $\alpha_r=1$ 
and $\alpha_v=1$ will get rid of the tail almost completely. 
However, the value of $f$ is quite low for low-mass galaxies 
and is lower when a more restrictive limit is applied, 
indicating that many of the interlopers are located in the outer 
parts of halos. The fact that a substantial fraction of 
low-mass galaxies are located beyond $R_{\rm vir}$
and have relative velocities larger than $V_{\rm vir}$
is because the groups identified by the group finder are 
usually non-spherical, particularly in high density 
regions. Note that the $\rm\log(M_{h, fit} / M_{h, t})$ 
distributions shown in Fig.\,\ref{delta_hm_hist} are weighted 
by the number of galaxies in halos, so that the extended tails 
in the distributions are dominated by a small number of 
systems in high density regions where the contamination by 
interlopers is severe. In any case, for investigations  
where purity of member galaxies is crucial, one should 
adopt restrictive limits on $\Delta {\rm r_p}$ and 
$\Delta v$ to reduce the contamination by interlopers. 

\section{The application to the zCOSMOS-bright sample}%
\label{sec:the_application_to_the_zcosmos_sample}
	
The zCOSMOS-bright is a spectroscopic galaxy survey obtained 
with the ESO VLT \citep{lillyZCOSMOSLargeVLT2007, Lilly:2009fb}.
It contains about 20,000 galaxies with $15.0 \leq I_{\rm AB} \leq 22.5$ in 
an area of about 1.7deg$^2$ in the COSMOS field and in  
the redshift range $0.1\lesssim z \lesssim 1.2$.
The redshift completeness, defined as the product of 
the redshift sampling rate and the redshift success rate
\citep{Knobel:2012je}, is $\sim 48\%$ in the full zCOSMOS-bright area 
and $\sim 56\%$ in the central region.
As discussed in \citet{de_la_torre_zcosmos-bright_2011},
the sampling effects for zCOSMOS can be modeled as a function of the 
right ascension (RA) and redshift.
As an application of our group 
finding pipeline, we will identify galaxy groups in the central region of
the COSMOS area using both spectroscopic and photometric 
galaxies at $0.1 \leq z \leq 1.0$.
	
\subsection{Tests with zCOSMOS-bright mock samples}%
\label{sub:tests_with_zcosmos_mock_samples}

\begin{figure*}
	\includegraphics[width=\linewidth]{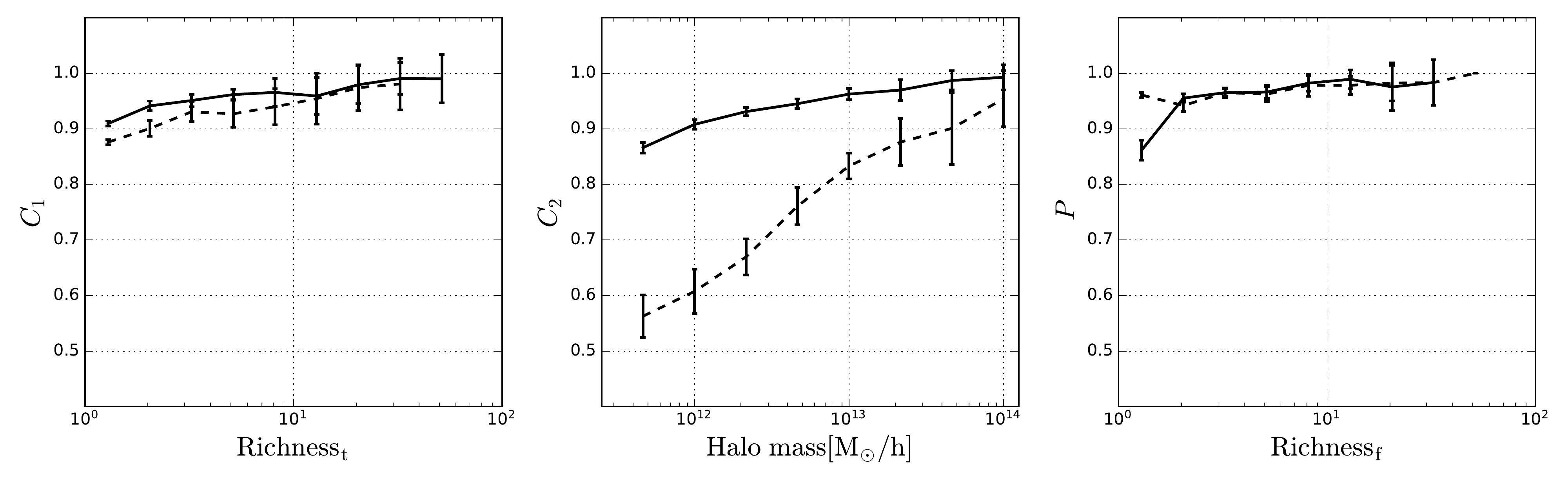}
	\caption{Performance of our group finder on the zCOSMOS-bright mock catalog
        in terms of $C_1$, $C_2$ and $P$ (see Table~\ref{symbols} for definitions).
        The dashed lines are for the spectroscopic only sample and the solid 
        lines are the performance including photometric data. Error bars show 
        the standard deviations among 20 different mock samples.}
	\label{group_performance_zcosmos}
\end{figure*}

\begin{figure}
	\includegraphics[width=\linewidth]{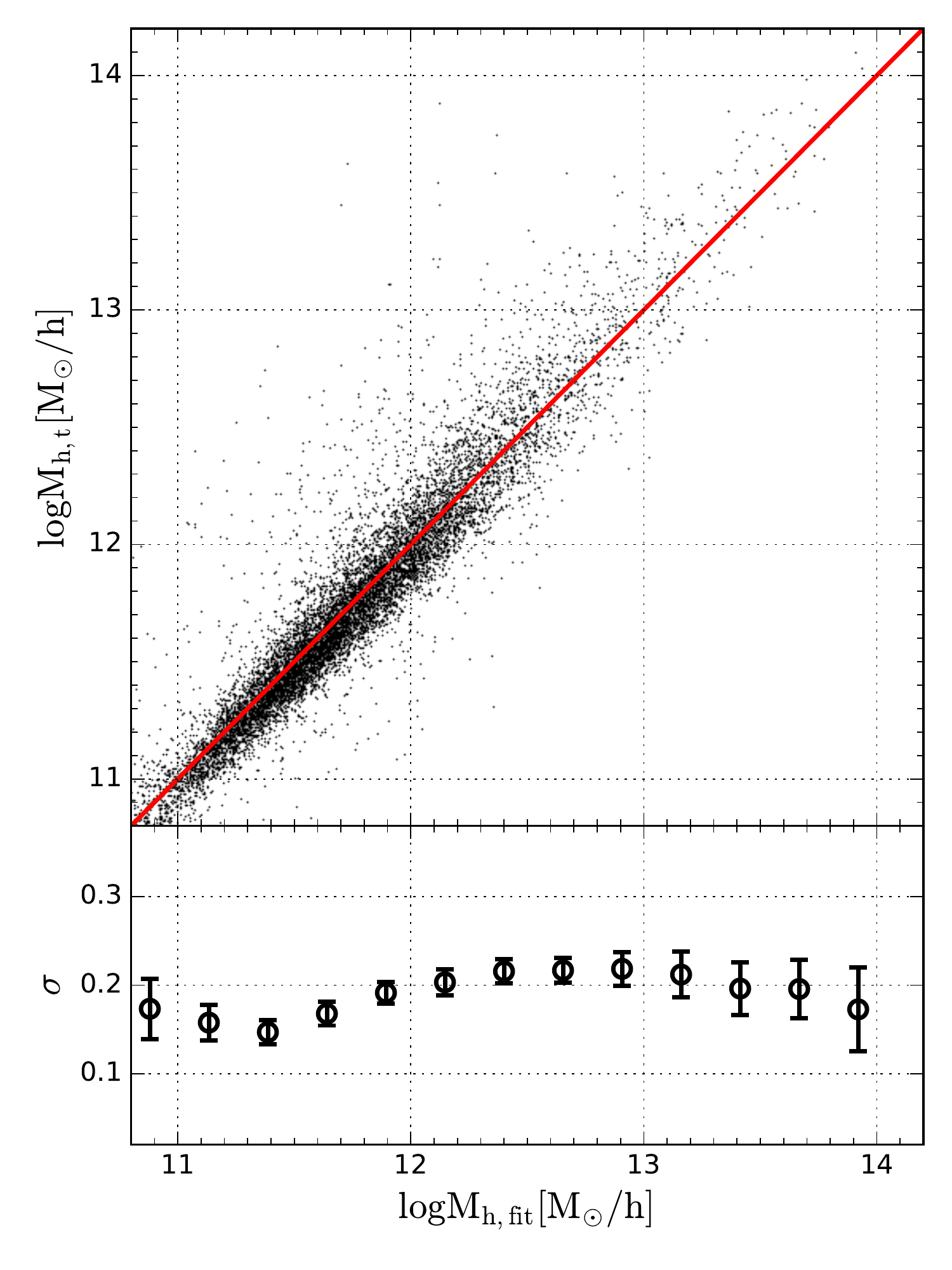}
	\caption{Performance of the group finder on halo mass for  
        zCOSMOS-bright mock catalogs, shown as the relationship between 
        the true halo mass, $M_{\rm h,t}$, and the predicted halo mass, 
        $M_{\rm h,fit}$. The standard deviations of true halo halo mass for a given 
        predicted mass are shown in the lower panel as circles, with 
        error bars representing the variances among 20 mock samples.}
	\label{halomass_performance_zcosmos}
\end{figure}

\begin{figure*}
	\includegraphics[width=\linewidth]{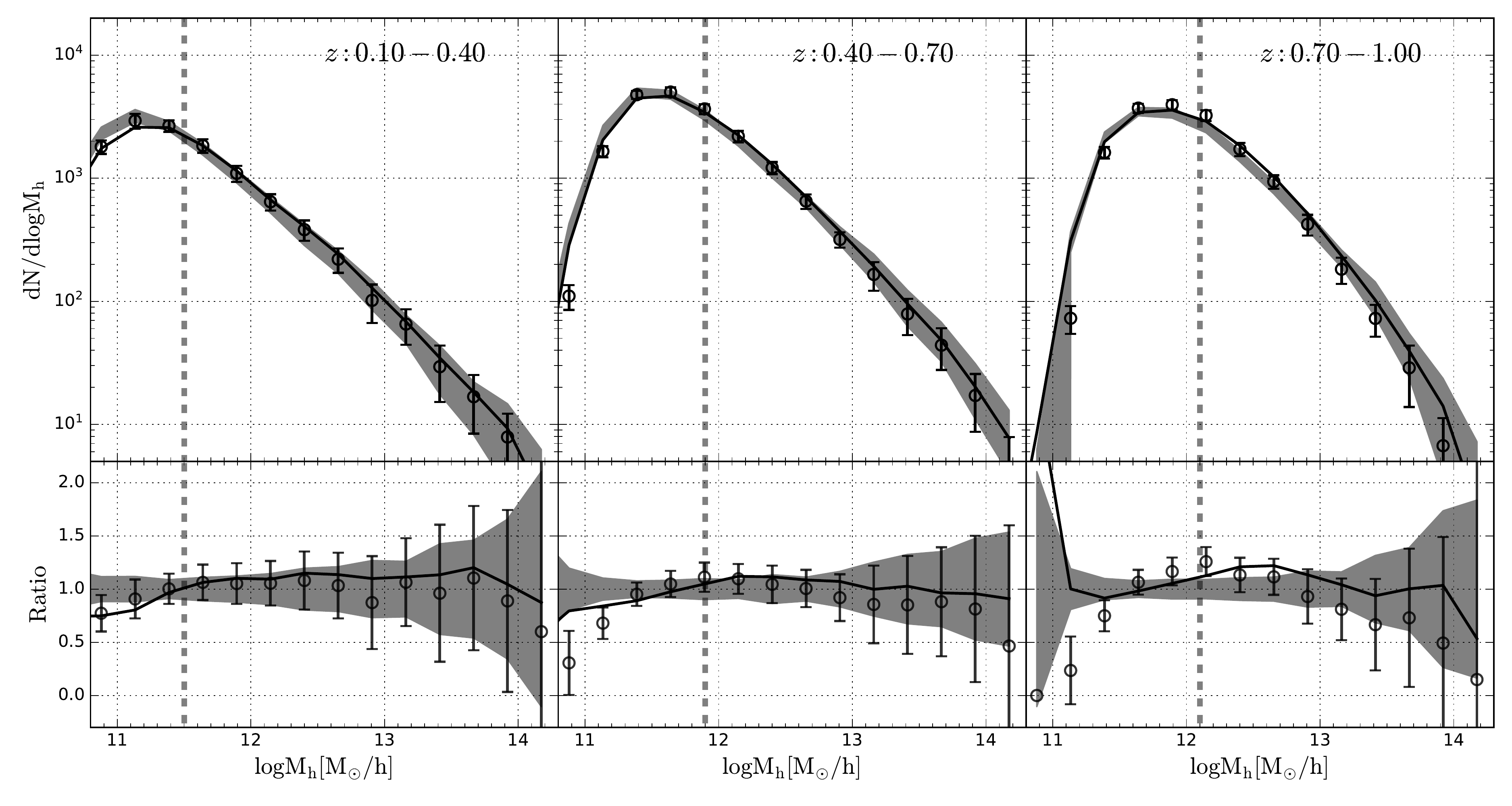}
    \caption{Halo mass functions for identified groups in 
    zCOSMOS-bright mock samples. 
    The Grey shaded regions are the ranges covered by the halo mass distribution 
    by the 20 mock samples. The data points with error bars are for 
    identified groups with estimated halo masses. The solid lines are for 
    $\rm M_{h, samp}$. The vertical dashed lines are the mass limits 
    reached by the catalog. The lower panels show the same functions
    but normalized by the mean of the 20 mock samples.}
	\label{halomassfunction_zcosmos}
\end{figure*}

\begin{table}
	\centering
	\caption{Optimal parameters of friend-of-friend group finder in the central region for zCOSMOS-bright survey.}
	\label{para_finder_zcosmos}
	\begin{tabular}{c|c|c|c}
		\hline
		Parameters & $b$  & $l\_{\rm max}$(Mpc/h) & $R$   \\
		\hline\hline
		Values     & 0.08 & 0.30                  & 17.00 \\
		\hline
	\end{tabular}
\end{table}

To quantify the performance of our group finding pipeline on the zCOSMOS-bright like 
surveys, we constructed 20 different mock catalogs to mimic the selection effects 
and incompleteness for the central region of the real zCOSMOS-bright survey 
in the redshift range of $0.1 \le z \le 1.0$ \citep{mengMeasuringGalaxyAbundance2020}.
	
The group level performance of our group finder for the zCOSMOS-bright mock samples,
which uses the optimal parameters listed in Table\,\ref{para_finder_zcosmos},
is shown in Fig.\,\ref{group_performance_zcosmos}. The dashed lines are based on 
spectroscopic-only galaxies, while solid lines use both spectroscopic and 
photometric galaxies. Similar to the results presented above, our group finder 
performs well in terms of both $C_1$ and $P$ (both $\gtrsim 90\%$). 
A large deficit in the $C_2$ index is observed when only spectroscopic galaxies 
are used, especially for low mass halos, but the inclusion of 
photometric galaxies improves the performance dramatically.

We also estimate the halo masses using the RFR as described above,
and the performance
is shown in Fig.\,\ref{halomass_performance_zcosmos}. 
Over the entire mass range from 
$\rm\sim 10^{11}M_{\odot}/h$ to $\rm\sim 10^{14}M_{\odot}/h$, 
the standard deviation of the estimated halo mass is about 0.2 dex. 
The estimated halo mass functions are shown in 
Fig.\,\ref{halomassfunction_zcosmos} as data points with error bars,
in comparison with those obtained directly from 20 mock samples
(gray regions). The black solid lines are the average distribution
function of $\rm M_{h, samp}$ among 30 random samples
obtained using equation\,(\ref{hm_samp}).
For comparison, the mass limit for completeness is indicated as 
vertical dashed line in each panel.
As one can see, the input halo mass functions
can be well recovered; the large scatter at the massive end 
among different mock samples reflects the level of the cosmic 
variance expected for a sample like zCOSMOS-bright. 

\subsection{The zCOSMOS-bright group catalog}
\label{zcosmos_data}

\begin{figure*}
	\includegraphics[width=\linewidth]{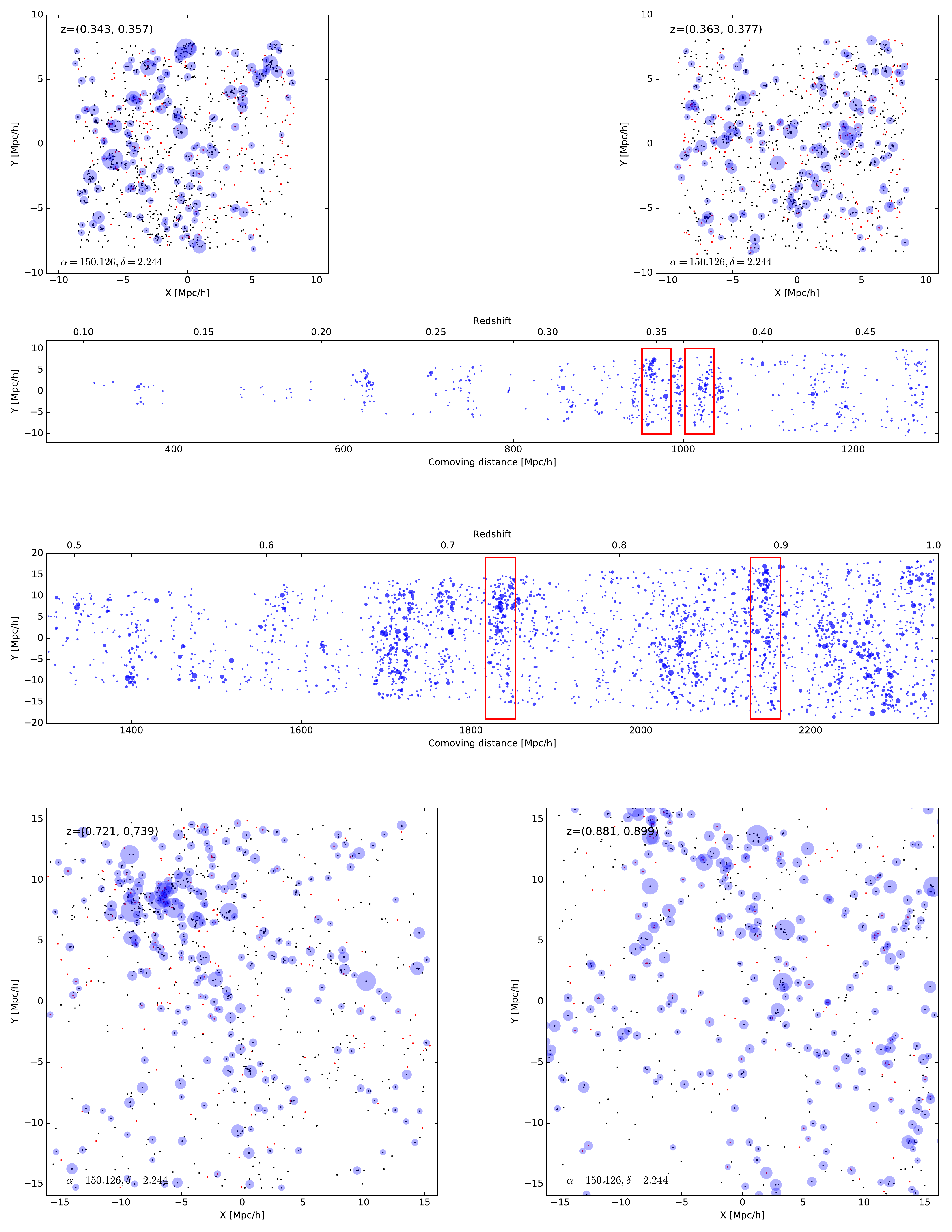}
	\caption{The projected distributions of galaxies and identified groups in 
	four redshift slices. The black dots are spectroscopic galaxies and the red dots 
	are photometric galaxies. The blue circles represent the galaxy groups 
	with $\rm M_h>10^{12}M_{\odot}/h$), with radius proportional 
	to the halo radius.}
	\label{region_plot}
\end{figure*}

\begin{figure*}
	\includegraphics[width=\linewidth]{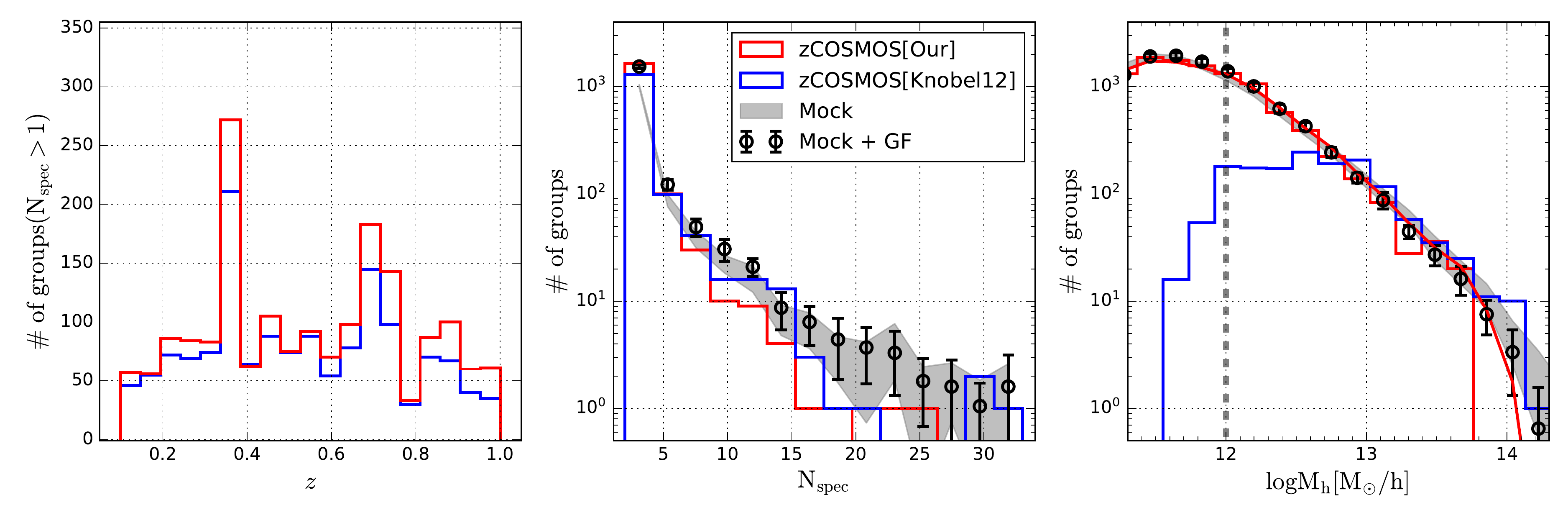}
    \caption{Comparison between our group catalog with that of 
    \citet{Knobel:2012je}. Left panel: redshift distribution; Middle panel: richness 
    distribution; Right panel: halo mass distribution. Our results 
    are shown by red histograms while those of \citet{Knobel:2012je} by blue 
    histograms. The red solid curve in the right panel is 
    the distribution of $\rm M_{h, samp}$ obtained from our catalog.
    Circles with error bars are the mean and variance obtained by applying 
    our group finder to the 20 zCOSMOS-bright mock catalogs, while   
    the grey regions cover the ranges obtained directly from the 20 
    mock catalogs. 
    The vertical dashed line in the right panel indicates  
    the completeness limit. Note that the halo masses are available 
    only for groups that contain at least 2 spectroscopic members
    in the catalog of \citet{Knobel:2012je}.}
	\label{compare_knobel12}
\end{figure*}

We have applied our group finder to zCOSMOS-bright galaxies 
at $0.1 \leq z \leq 1.0$ in the central region that covers $\sim $1 deg$^2$.
We also excluded unreliable redshift measurements tagged as 0, 1.1, 2.1 and 9.1 
\citep{Lilly:2009fb}. The final spectroscopic sample contains 11,489 galaxies. 
The photometric data used is adopted from the parent photometric sample, 
constructed from \citet{laigleCOSMOS2015CATALOGEXPLORING2016} by 
\citet{mengMeasuringGalaxyAbundance2020}. The spectroscopic groups are identified 
using the FoF group finder with optimal parameters calibrated by 
the mock samples (see Table\,\ref{para_finder_zcosmos}). 
Starting from the spectroscopic groups, we identify both 
{\it isolated centrals} and {\it group centrals} that are 
missed in the spectroscopic sample based on the 
parent photometric sample, using the RFC method described in
\S\,\ref{sub:supplementing_with_photometric_galaxies}. 
Finally, we calibrate the halo masses for 
the final group catalog using the RFR described in 
\S\,\ref{sub:assigning_halo_masses_to_groups}.

Fig.\,\ref{region_plot} shows the spatial distribution of the 
identified groups in the $(Y, Z)$ plane (the two middle panels)
where $Z$ is in the radial (redshift) direction, and $Y$ is one of the 
two directions perpendicular to $Z$. As illustrations, 
the four square panels in the upper and lower rows  
show the distribution in the $X$-$Y$ plane for groups  
in four redshift slices with $\Delta z=0.01(1 + z)$, as indicated 
by the four red rectangles. Only groups with 
$\rm M_h\ge 10^{12}M_{\odot}/h$ are plotted, and each 
of them is shown as a blue circle with radius proportional to its halo 
radius. For comparison, we also show spectroscopic galaxies 
as black points, and photometric galaxies as red points. 
We can see clearly, as expected, that galaxy groups trace the large-scale 
structure in the galaxy distribution, and that massive groups 
reside preferentially in high density regions. 

We plot the redshift ($z$) distribution of our identified groups 
in Fig.\,\ref{compare_knobel12}, in comparison  
with that obtained by \citet{Knobel:2012je}. Despite of the 
different methods used to identify galaxy groups, the two 
distributions match well with each other.
Fig.\,\ref{compare_knobel12} also shows the richness and halo 
mass distributions of our group catalog, again in comparison 
with those obtained from the catalog of \citet{Knobel:2012je}. 
Both group catalogs give a similar distribution in the richness of 
spectroscopic members. This is expected, as we are using a
similar method to identify groups in the spectroscopic sample. 
However, our catalog contains many more low-mass systems, 
because we include isolated systems and our halo mass 
estimator provides reliable mass estimates even for low-mass 
halos. There is also discrepancy between the two catalogs at the massive 
end, where our group catalog contains smaller number of groups. 
We believe that this owes to the galaxy number density re-calibration 
used by \citet{Knobel:2012je}, as described below. As a demonstration, 
the circles with error bars in Fig.\,\ref{compare_knobel12}
show the result obtained by applying our group finder to the 20 
zCOSMOS-bright mock catalogs, in comparison to that obtained directly 
from the mock catalogs, shown by the grey regions. The fact that these 
two results match well with each other indicates that our group finder is 
reliable. The discrepancy between our zCOSMOS-bright results and the mock 
results then suggests that the zCOSMOS-bright is not a fair sample, 
particularly for massive groups.

\citet{Knobel:2012je} published a galaxy group catalog based on the 
spectroscopic galaxies from zCOSMOS 20k, using the FoF group finding 
algorithm in a "multi-run scheme", and using photometric galaxies 
to make improvements on group membership and group center.
They calibrated their FoF parameters and halo mass estimator using mock catalogs 
that are scaled so that the average density distribution of galaxies 
matches that in the real sample. 
Thus, their results are, in a sense, corrected for cosmic variance. 
This may explain why their group mass function matches the 
expected mass function better at the massive end (see the right panel of 
Fig.\,\ref{compare_knobel12}).
In this paper, we decide to provide a group catalog 
that is based on the data itself, while leaving the correction for the 
cosmic variance to specific applications of the catalog. 
In addition, our group finding algorithm is different from that 
of \citet{Knobel:2012je} in the following aspects.
First, we use the state of the art random forest algorithm to incorporate 
photometric galaxies and to improve the completeness and purity of our 
group catalog. Second, we use a halo mass estimator, calibrated with 
realistic mock catalogs and the random forest method, so that 
we are able to provide accurate halo mass estimates for groups over a 
large mass range. 

\subsection{Catalog contents}

The group catalog constructed and the galaxy sample 
used for the construction are available
through \url{https://github.com/wkcosmology/zCOSMOS-bright_group_catalog}. 
The group catalog lists the properties of individual groups, 
while the galaxy sample provides information about 
individual galaxies as well as their links to groups. 
In what follows we explain the contents of these 
catalogs in more detail.

\subsection{The group catalog}%
\label{sub:the_group_catalog}

The following items are provided for individual groups.
\\
Column (1) \texttt{groupID}: a unique ID of each group in the
group catalog;
\\
Column (2) \texttt{cenID}: galaxy ID of the central galaxy of a
group;
\\
Column (3) \texttt{cenID2015}: central galaxy ID in \citet{laigleCOSMOS2015CATALOGEXPLORING2016};
\\
Column (4) \texttt{RA\_avg}: Right Ascension (J2000) of
the group center in degrees, defined as the average RA of 
member galaxies weighted by the stellar mass
\\
Column (5) \texttt{Dec\_avg}: Declination (J2000) of the
group center in degrees, defined as the average Dec of member 
galaxies weighted by the stellar mass
\\
Column (6) \texttt{z\_avg}: redshift of the group,
defined as the average redshift of member galaxies with spectroscopic 
redshift weighted by the stellar mass
\\
Column (7) \texttt{HaloMass}: 10-based logarithm 
of the halo mass of a group in units of $\rm M_{\odot}$;
\\
Column (8) \texttt{GroupTag}: 0 for groups with only spectroscopic members, 
1 for groups with photometric central and spectroscopic member, 
and 2 for groups with only one photometric member;
Column (9) \texttt{Richness}: number of member galaxies in a group;
\\

\subsection{The galaxy catalog}%
\label{sub:the_galaxy_catalog}

The following items are provided for individual galaxies
\\		
Column (1) \texttt{ID}: unique ID of galaxies, which  
can be used to match galaxies across the galaxy and group catalogs;
\\
Column (2) \texttt{surveyID}: ID of galaxies from the original
survey data release. This can be used to match
galaxies across our catalogs and the original
survey data release;
\\
Column (3) \texttt{ID2015}: galaxy id in 
\citet{laigleCOSMOS2015CATALOGEXPLORING2016};
\\
Column (4) \texttt{groupID}: ID of the group of which a galaxy
is a member;
\\
Column (5) \texttt{RA}: right ascension (J2000) in degrees;
\\
Column (6) \texttt{Dec}: declination (J2000) in degrees;
\\
Column (7) \texttt{z}: redshift
\\
Column (8) \texttt{StellarMass}: 10-based logarithm of the galaxy
in units of $\rm M_{\odot}$;
\\
Column (9) \texttt{tag}: 1 for central, 0 for satellite;
\\
Column (10) \texttt{CC}: redshift confidence class, $-1$ for photometric 
redshift, others see \citet{lillyZCOSMOSLargeVLT2007}.

\section{Summary}%
\label{sec:summary}

In this paper, we have developed a group finder that is 
suitable for identifying galaxy groups from incomplete
redshift samples combined with photometric data. A machine
learning method is
adopted to assign halo masses to identified groups.  
To test the impact of redshift sampling effects, we have 
constructed realistic mock samples with different redshift 
sampling schemes and applied our group finder to them.  
Our main results are summarized as follows.
\begin{enumerate}
    \item
        We find that our modified version of the FoF group finder 
        based on a local, incompleteness-corrected  
        linking-length can identify most of the galaxy systems correctly 
        from an incomplete spectroscopic sample 
        (Fig.\,\ref{group_performance_pfs}), even with a sampling 
        rate that is as low as 55\% and is spatially in-homogeneous.
    \item
        We find that an incomplete redshift sampling can cause the loss of 
        galaxy groups from a spectroscopic sample. 
        For random sampling cases, many of the low-mass groups are lost 
        although the massive ones can still be identified due to their high 
        richness. However, with realistic fiber assignments, such as  
        the one to be adopted by the up-coming PFS galaxy survey, 
        massive galaxy systems can also be missed because of the 
        lower sampling rates in higher density regions caused by fiber 
        collisions (Fig.\,\ref{group_performance_pfs}).
    \item
        With the use of the state-of-the-art random forest algorithm, 
        we find that it is possible to retrieve most of the lost 
        groups using a combination of spectroscopic and 
        photometric data. The final completeness and purity that can be 
        achieved can reach to $\gtrsim 85\%$ (Fig.\,\ref{group_performance_pfs})
        even for a sampling rate as low as 55\% and for an in-homogeneous sampling.
    \item
        We calibrate the host halo mass for identified galaxy groups with 
        the random forest regressor algorithm. We find that the estimated 
        halo masses are un-biased relative to the true masses, with an 
        uncertainty of about 0.15 -- 0.25 dex over a wide range of
        halo masses
        (Fig.\,\ref{halomass_performance_pfs}). The estimated halo mass distribution 
        matches the input mass function well after the statistical bias 
        caused by the mass uncertainty is taken into account. 
    \item
        We find that the conditional stellar mass functions of galaxies 
        in halos of different masses can be well recovered from the identified 
        groups with estimated halo masses (Fig.\,\ref{csmf}). 
    \item 
        We find that the groups identified by our group finder 
        provide an accurate link between individual 
        galaxies and the masses of their host halos (Fig.\,\ref{delta_hm_hist}).
        Although there are some interlopers with high $\rm\log(M_{h, fit}/M_{h, t})$,
        we have shown that these outliers can be eliminated by cutting out 
        members in the outer parts of groups.
    \item
        We have applied our group finding algorithm to the 
        zCOSMOS-bright spectroscopic 
        redshift survey and constructed a new catalog of galaxy groups in 
        $0.1\le z\le 1.0$. Our tests using mock catalogs show that 
        most of the galaxy groups are identified
        correctly (Fig.\,\ref{group_performance_zcosmos}) with  
        reliable halo masses (Fig.\,\ref{halomass_performance_zcosmos}).
        Compared with the previous group catalog selected from 
        the zCOSMOS-bright survey, our catalog is more complete, extending
        the halo mass range to much lower masses. 
        Our halo mass estimates
        are reliable over the entire mass range covered by our catalog, 
        as shown by our tests based on realistic mock catalogs.
\end{enumerate}

Identifying galaxy groups from redshift surveys of galaxies
plays an important role in connecting galaxies with the underlying 
dark matter distribution. Our results demonstrate clearly that 
such investigations can also be carried out for current and future 
high-$z$ spectroscopic surveys. This opens a new avenue to  
connect galaxies to their dark matter halos at high $z$, thereby 
to study galaxy evolution in different environments.    
Furthermore, the success of our method to construct 
highly complete 
group samples covering large halo mass ranges demonstrates 
that galaxy groups properly identified at high $z$ can be 
used to represent the dark halo population in the early universe.   
One can thus use them to reconstruct the cosmic density field
and to study the large-scale structure in the early universe,  
as was done in low $z$ 
\citep[][]{Wang_Mo_Jing_Guo_vandenBosch_Yang_2009}. One 
can also use the galaxy groups as tracers to investigate the
properties of dark matter halos 
at high $z$ through, e.g., their gravitational lensing 
effects and Sunyaev–Zel'dovich effects.

\section*{Acknowledgements}

This work is supported by the National Key R\&D Program of China
(grant No. 2018YFA0404502, 2018YFA0404503), and the National Science Foundation of
China (grant Nos. 11821303, 11973030, 11673015, 11733004,
11761131004, 11761141012). Part of our analysis is 
based on data products from observations made with ESO Telescopes at 
the La Silla Paranal Observatory under ESO programme ID 179.A-2005 and 
on data products produced by TERAPIX and the Cambridge Astronomy Survey 
Unit on behalf of the UltraVISTA consortium. Kai Wang and Yangyao Chen
gratefully acknowledge the financial support from China Scholarship Council.

\section*{Data availability}

The data underlying this article will be shared on reasonable request to
the corresponding author. The zCOSMOS-bright group catalog are available at
\url{https://github.com/wkcosmology/zCOSMOS-bright_group_catalog}. 


\appendix

\section{The importance of different features used for halo mass estimate}%
\label{sec:feature_importance_for_halo_mass_estimation}

\begin{figure*}
    \centering
    \includegraphics[width=1\linewidth]{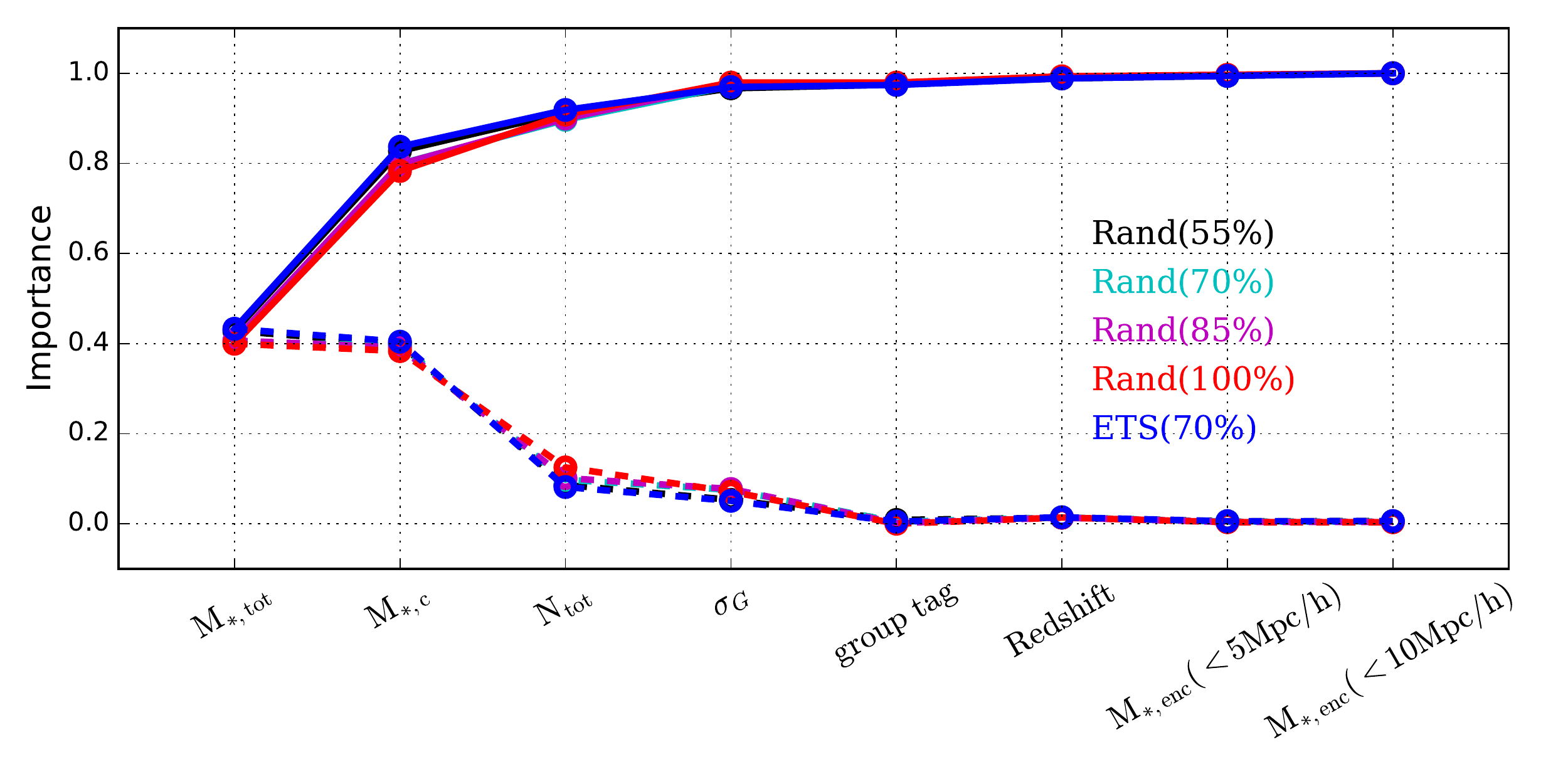}
    \caption{Feature importance (dashed lines) and the corresponding cumulative  
    distribution (solid lines) for different sampling cases.}%
    \label{fig:figure/feature_imptances}
\end{figure*}

We employ the RFR to predict the halo mass for galaxy 
groups (\S\ref{sub:assigning_halo_masses_to_groups}), using several group 
properties as input features. RFR also provides a way to quantify the 
contribution of each individual feature to the prediction in terms of 
feature importance. Recall that the random forest is assembled by many 
decision trees, each of which is constructed by iteratively bi-partitioning 
the sample into left and right children with one feature, and 
each bi-partition is to minimize a certain goal function
(like Gini impurity for RFC, and the mean squared error for RFR). 
Heuristically, if a feature is always chosen to bi-partition the 
tree and the bi-partitions can dramatically decrease the goal 
function, this feature must be important in predicting the target value. 
The importance of feature-$i$ can thus be calculated for a decision tree 
though
\begin{equation}
    {\rm Imp}_i = {\sum_{j:{\rm nodes~splitted~according~to~feature-}i} 
    {\rm \Delta MSE}_j\over \sum_{j: {\rm all~nodes}} {\rm \Delta MSE}_j}
\end{equation}
where the summation $j$ is for the internal nodes. The quantity ${\rm \Delta MSE}_j$ 
is the MSE decrement for each $j$-th internal node, defined as
\begin{align}
    {\rm \Delta MSE}_j &= \sum_l^{|\mathcal{D}_j|}(y_l - \bar y_j)^2\\
                       &- \sum_l^{|\mathcal{D}_{j, \rm L}|}(y_l - \bar y_{j, \rm L})^2 - 
                       \sum_l^{|\mathcal{D}_{j, \rm R}|}(y_l - \bar y_{j, \rm R})^2
\end{align}
where $\bar y_j$ is the target mean of data points in node $j$;
$\bar y_{j, \rm L}$ and $\bar y_{j, \rm R}$ are the target means for 
the left and right children, respectively;
$|\mathcal{D}_j|$, $|\mathcal{D}_{j, \rm L}|$ and $|\mathcal{D}_{j, \rm R}|$ 
are the numbers of data points in node $j$ and in its left and right children, 
respectively. Fig.\,\ref{fig:figure/feature_imptances} shows the importance of 
different features adopted in the main text to 
determine the halo mass, with the total 
importance normalized to unity. As one can see, the total stellar mass, 
central stellar mass, richness and velocity dispersion are the four 
features dominating the contribution, while other features 
contribute little.

\bibliographystyle{mnras}
\bibliography{bibtex.bib}
	
\bsp	
\label{lastpage}
\end{document}